\documentclass[aps,pra,showpacs,amssymb,superscriptaddress,nofootinbib,twocolumn]{revtex4-2}

\usepackage{overpic,color}
\usepackage{amsmath,amsfonts,amsthm,amssymb}
\usepackage{braket}
\usepackage{setspace}
\usepackage{amsmath}
\usepackage{appendix}
\usepackage[utf8]{inputenc}
\usepackage{bbm}
\usepackage{bm}
\usepackage{amsbsy}
\usepackage{dsfont} 
\usepackage{graphicx} 
\usepackage{epsfig}
\usepackage{epstopdf}
\usepackage{dsfont}
\usepackage{color}
\usepackage{soul}
\usepackage[bookmarks=true, colorlinks]{hyperref}
\usepackage{upgreek}
\usepackage[figure,table]{hypcap}
\usepackage{enumerate}
\usepackage{natbib}

\makeatletter
\newcommand\org@hypertarget{}
\let\org@hypertarget\hypertarget
\renewcommand\hypertarget[2]{
  \Hy@raisedlink{\org@hypertarget{#1}{}}#2%
  }
\makeatother

\hypersetup{
	bookmarksnumbered,
	pdfstartview={FitH},
	citecolor={darkgreen},linkcolor={darkred},
	urlcolor={darkblue},
	pdfpagemode={UseOutlines}}
\definecolor{darkgreen}{RGB}{50,190,50}
\definecolor{darkblue}{RGB}{0,0,190}
\definecolor{darkred}{RGB}{238,0,0}

\newcommand{\Ca}{$^{40}\text{Ca}^+$}

\begin{document}

\title{Towards a deterministic interface between trapped-ion qubits and travelling photons}
\vspace{20mm}

\author{J. Schupp}
 \affiliation{Institut f\"ur Quantenoptik und Quanteninformation,\\
 	\"Osterreichische Akademie der Wissenschaften, Technikerstr. 21A, 6020 Innsbruck,
 	Austria}
 \affiliation{
 	Institut f\"ur Experimentalphysik, Universit\"at Innsbruck,
 	Technikerstr. 25, 6020 Innsbruck, Austria}
 	
\author{V. Krcmarsky}
\affiliation{Institut f\"ur Quantenoptik und Quanteninformation,\\
	\"Osterreichische Akademie der Wissenschaften, Technikerstr. 21A, 6020 Innsbruck,
	Austria}
\affiliation{
	Institut f\"ur Experimentalphysik, Universit\"at Innsbruck,
	Technikerstr. 25, 6020 Innsbruck, Austria}

\author{V. Krutyanskiy}
\affiliation{Institut f\"ur Quantenoptik und Quanteninformation,\\
	\"Osterreichische Akademie der Wissenschaften, Technikerstr. 21A, 6020 Innsbruck,
	Austria}
\affiliation{
	Institut f\"ur Experimentalphysik, Universit\"at Innsbruck,
	Technikerstr. 25, 6020 Innsbruck, Austria}

\author{M. Meraner}
\affiliation{Institut f\"ur Quantenoptik und Quanteninformation,\\
	\"Osterreichische Akademie der Wissenschaften, Technikerstr. 21A, 6020 Innsbruck,
	Austria}
\affiliation{
	Institut f\"ur Experimentalphysik, Universit\"at Innsbruck,
	Technikerstr. 25, 6020 Innsbruck, Austria}

\author{T. E. Northup}
\affiliation{
	Institut f\"ur Experimentalphysik, Universit\"at Innsbruck,
	Technikerstr. 25, 6020 Innsbruck, Austria}

\author{B. P. Lanyon}
\affiliation{Institut f\"ur Quantenoptik und Quanteninformation,\\
	\"Osterreichische Akademie der Wissenschaften, Technikerstr. 21A, 6020 Innsbruck,
	Austria}
	\affiliation{
	Institut f\"ur Experimentalphysik, Universit\"at Innsbruck,
	Technikerstr. 25, 6020 Innsbruck, Austria}

\begin{abstract}
\vspace{5mm}
Experimental results are presented on the efficiency limits for a quantum interface between a matter-based qubit and a photonic qubit. Using a trapped ion in an optical cavity, we obtain a single ion-entangled photon at the cavity output with a probability of 0.69(3). The performance of our system is shown to saturate the upper limit to photon-collection probability from a quantum emitter in a cavity, set by the emitter's electronic structure and by the cavity parameters. The probability for generating and detecting the ion-entangled fiber-coupled photon is 0.462(3), a five-fold increase over the previous best performance. Finally, the generation and detection of up to 15 sequential polarised photons demonstrates the ability of a trapped ion to serve as a multi-photon source.
The comparison between measured probabilities and predicted bounds is relevant for quantum emitters beyond trapped ions, in particular, for the design of future systems optimising photon collection from, and absorption in, quantum matter.
\end{abstract}

\maketitle

The development of interfaces between travelling photons and quantum matter is a key requirement for emerging quantum technologies, allowing for single photon sources \cite{Solomon2013} and the transfer, storage and redistribution of quantum information \cite{Heshami2016}.
If the quantum matter is a register of qubits with quantum logic capabilities, then more powerful applications become possible, such as scalable quantum computing \cite{Monroe2014} and arbitrary-distance quantum networks \cite{Kimble2008, Wehner2018} for secure communication \cite{Gisin2002}, distributed quantum sensing \cite{Sekatski2020} and enhanced timekeeping \cite{Komar2014}. 

The efficiency of the interface---the probability with which photons can be collected from or absorbed by matter qubits---is a key parameter to optimise for the practical realisation of the aforementioned  quantum technologies. 
Paths to deterministic interfaces include collective effects in interacting particle ensembles \cite{Duan2001} or optical resonators that exploit vacuum-modification effects to enhance emission into and absorption from a desired optical mode, e.g., Fabry-Pérot cavities \cite{Reiserer2015}, micro-resonators \cite{Junge2013, Wang2019a} and nano-photonic waveguides \cite{Tiecke2014, Hummel2019}. 

Efficient photonic interfaces for trapped-ion qubits are desirable: the deterministic, near fault-tolerant and multi-qubit quantum information processing capabilities of the ion-trap platform \cite{Bruzewicz2019, Friis2018, Zhang2017, Haffner2005} could enable arbitrary-distance quantum networking via the repeater approach \cite{Sangouard2009, Munro2015} and scalable quantum computing \cite{Monroe2014}. 
Furthermore, trapped ions are amongst the most precise sensors \cite{Baumgart2016} and clocks \cite{Brewer2019} ever developed and could be used to construct distributed networks of quantum-limited sensors.

Combining trapped ions with high numerical aperture lenses is a powerful approach to connecting the quantum states of  trapped-ion qubits to travelling photons \cite{Moehring2007, Hucul2015}. 
Single photons on demand, emitted by and entangled with a trapped-ion qubit, have been generated and detected with a total probability of 0.024 after collection via an in-vacuum lens \cite{Stephenson2020}.
The first use of an optical cavity to enhance the collection of ion-entangled photons achieved a combined probability for generation and detection -- in the following simply called detected probability -- of 0.057(2) \cite{Stute2012}.
In Refs. \cite{Krutyanskiy2019, efficiency50km} a detected probability of 0.08(1) was achieved using a similar cavity design as in Ref. \cite{Stute2012}. 
Ion-entangled photons were recently collected using a microscopic fiber cavity, with a detected probability of $2.5\times 10^{-3}$ \cite{Kobel2021}. 
The first key result of this manuscript is a detected probability of 0.462(3) for ion-entangled photon.
This result is achieved using the macroscopic optical cavity of Ref. \cite{Krutyanskiy2019}.

The second key result of this manuscript is that the probability with which an ion-entangled photon is obtained at the cavity output could not be significantly higher without changing properties of our cavity or quantum emitter. Specifically, the performance of our system saturates recently derived theoretical limits on the photon collection efficiency from an emitter-cavity system \cite{Goto2019}, set only by the cavity waist, losses, photon wavelength and electronic structure of the quantum emitter. These results provide clear paths for future improvements in the efficiency of photon-emitter interfaces.  

The paper layout is as follows:
Sec. \ref{sec:model} summarises recently developed analytical expressions for the limits to the probability of photon collection for a quantum emitter using a cavity.
The employed model is general enough to apply to a broad range of physical systems and photon generation schemes.
Sec. \ref{sec:exp_det} presents experimental details for our ion-cavity system, which by careful design achieves close to the optimal compromise between photon emission probability into the cavity mode and exit probability through the desired output mirror.  
Secs. \ref{sec:exp_res} and \ref{sec:ent} present experimental results on the generation of single photons not entangled with the ion and single photons entangled with the ion, respectively.  
Sec. \ref{sec:mult_phot} presents experimental results on the generation of multiple sequential photons. 
In Sec. \ref{sec:improvements}, paths are identified to achieve efficiencies of over $0.90$ in future systems.
Sec. \ref{sec:conclusion} concludes the paper with a summary and outlook.

\section{Basic model}
\label{sec:model}

\begin{figure}[t]
	\vspace{0mm}
	\begin{center}
		\includegraphics[width=\columnwidth]{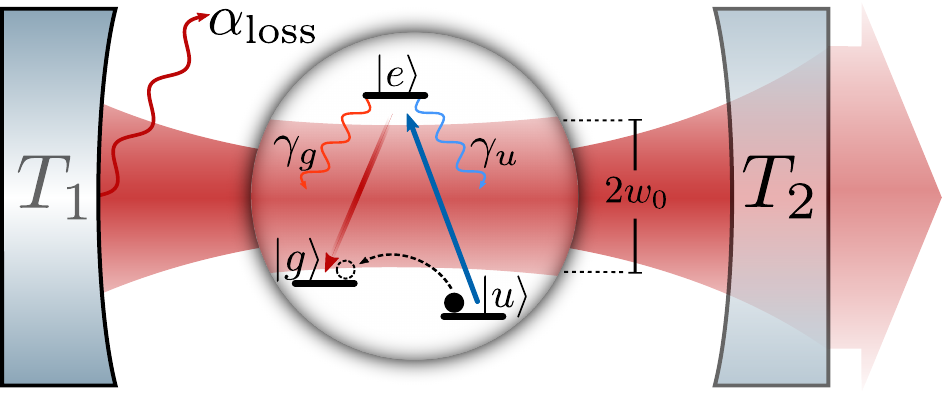}
		\vspace{0mm}
		\caption{
			\textbf{Limits to photon collection: a quantum emitter in a cavity.} 
			A three-level emitter in the waist $w_0$ of a vacuum mode of a cavity formed by two mirrors with transmission values $T_1$ and $T_2$.
			The unwanted cavity loss per round-trip $\alpha_{loss}$ includes $T_1$.
			The $\ket u \leftrightarrow \ket e$ transition is driven by an external field (blue arrow).  The $\ket g \leftrightarrow \ket e$ transition is coupled to the cavity (red arrow). 
			For the generation of photons, the system is initialised into the state $\ket{u,0}$, where $\ket 0$ is the vacuum Fock state of the cavity.
			Emission of a photon into the cavity leaves the system in $\ket{g,1}$.
			Spontaneous decay from $\ket{e}$ to $\ket{g}$ with decay constant $\gamma_g$ ends the attempt in failure, whereas decay from $\ket e$ to $\ket{u}$ with decay constant $\gamma_u$ allows for subsequent cavity photon generation (reexcitation processes).
			The maximum probability for a cavity photon to be emitted into the cavity and transmitted to the free-space output mode to the right of the cavity is given by Eq. \eqref{eq:PSopt} for $T_2^{opt}$ and by Eq. \eqref{eq:PS} for an arbitrary $T_2$.
		}
		\label{fig:model}
		\vspace{-6mm}
	\end{center}
\end{figure}

What is the maximum probability with which a photon can be collected from a quantum emitter? 
While ideal optical resonators, if they existed, could achieve unit probability, realistic ones do not due to unwanted photon loss mechanisms, such as scattering or absorption in the resonator mirrors \cite{Law1997, Kuhn2010, Vasilev2010}. 
Recently, expressions have been derived for the maximum photon collection probability 
for a single quantum emitter using a resonator with nonzero probability of unwanted photon loss \cite{Goto2019}. 
The model considered consists of a three-level quantum emitter with two ground states, $\ket u$ and $\ket g$, and one excited state $\ket e$.
The emitter is located at the position of the waist $w_0$ of a cavity formed by two mirrors with transmission values $T_1$ and $T_2$ (Fig. \ref{fig:model}).
The effective mode area of the cavity is ${A}_{e\!f\!f}=w_0^2 \pi /4$.
The $\ket u \leftrightarrow \ket e$ transition, with spontaneous decay rate $\gamma_u$, is addressed by an external drive field.
The $\ket e \leftrightarrow \ket g$ transition, with free-space spontaneous decay rate $\gamma_g$ and absorption cross-section $\sigma = 3\lambda^2/(2\pi)$ at wavelength $\lambda$, is addressed by a vacuum mode of the cavity.
All photon loss mechanisms of the cavity except for \textcolor{black}{the transmission} $T_2$ are captured by $\alpha_{loss}$, the unwanted cavity loss probability per round-trip.
The cooperativity is defined as $C = \frac{1}{\widetilde{A}_{e\!f\!f}(\alpha_{loss}+T_2)}\frac{\gamma_g}{\gamma}$ 
with $\widetilde{A}_{e\!f\!f}=A_{e\!f\!f}/\sigma$ and $\gamma$ the total spontaneous decay rate of $\ket e$. 
Here $\gamma = \gamma_g + \gamma_u$, but as we will see later, $\gamma$ may contain additional decay channels.

The output mode of the cavity is the free-space spatial mode that couples to the cavity via the mirror with transmission $T_2$. 
The probability for obtaining a photon in the output mode is $P_{S} = P_{in}P_{esc}$, where $P_{in}$ is the probability for a photon being emitted into the cavity and $P_{esc} = \frac{T_2}{T_2+\alpha_{loss}}$ is the probability for subsequent escape of the photon into the output mode.
In Ref. \cite{Goto2019} the authors show that $P_S$ is bounded via $P_S\le P^{bound}_S$ with
\begin{equation}
	P^{bound}_S = \left(\frac{T_2}{T_2+\alpha_{loss}}\right ) \left( \frac{2C}{1+2C}\right ) \sum_{j=0}^{\infty} \left( \frac{r_{u}}{1+2C}\right )^j,
	\label{eq:PS}
\end{equation}
where $r_u = \gamma_u/\gamma$ is the branching ratio for spontaneous decay on the transition $\ket e \rightarrow \ket u$.
The terms with $j>0$ are contributions to $P_{in}$ due to processes in which the emitter undergoes spontaneous decay from $\ket e$ to the initial ground state $\ket u$ before a cavity photon is generated.
These reexcitation events lead to cavity output photons that are not transform-limited, that is, the summation over $j$ describes a mixture of wave packets in the temporal domain. 
The extent to which this is detrimental depends on what the photons are to be used for. 
While reexcitation has no observable effect on the fidelity of our  emitter-photon entanglement (Sec. \ref{sec:ent}), it will reduce the indistinguishability of the photons, as we have recently studied \cite{Meraner2020}. 
We define $P^{pure}_S$ as the photon collection probability without reexcitation events, which is calculated by setting $j=0$ and is revisited in Sec. \ref{sec:improvements}.
In theory, $P^{bound}_S$ can be achieved via one of several different drive schemes, e.g., resonant excitation, off-resonant Raman scattering, or vacuum-stimulated Raman adiabatic passage \cite{Goto2019}.

Eq. \eqref{eq:PS} is maximised for the optimal output mirror transmission  
$T_2^{opt}=\alpha_{loss} \sqrt{1+\beta\frac 1{\alpha_{loss}}\frac 2{\widetilde{A}_{e\!f\!f}}}$ \cite{Goto2019}  
with $\beta=\frac{\gamma_g}{\gamma-\gamma_u}$, yielding the upper bound
\begin{equation}
	P^{opt}_{S} = 1-\frac 2{1+\sqrt{1+\beta\frac 1{\alpha_{loss}}\frac 2{\widetilde{A}_{e\!f\!f}}}}.
	\label{eq:PSopt}
\end{equation}
The existence of an optimal output mirror transmission can be understood by considering that there is a tradeoff relation in Eq. \eqref{eq:PS} between the probability for photon emission into the cavity ($P_{in}$) and the escape probability ($P_{esc}$) with respect to $T_2$: 
while $T_2 > T_2^{opt}$ would increase the probability for a photon already inside the cavity to be transmitted to the output mode, it would reduce the probability of the photon being emitted into the cavity in the first place, and the opposite would be true for $T_2 < T_2^{opt}$. 

In our experiment we have $\lambda = 854$ nm, $w_0 = 12.31(8)~\upmu$m and $\alpha_{loss}=26(4)$ ppm, and we have the following imperfections.
First, the relevant excited state $\ket e$ of our quantum emitter has decay channels to ground states other than $\ket g$ and $\ket u$, leading to $\beta = 0.70$
and resulting in $P^{exp.opt}_S = 0.84(1)$.
Second, our $T_2 = 90(4)$ ppm \textcolor{black}{is smaller than $T_2^{opt}$}, so the maximum achievable $P_S$ has to be calculated using Eq. \eqref{eq:PS}, yielding ${P}_{S} \le 0.75(3)$.
A more detailed discussion about the effects of a different value for $T_2$ in our setup can be found in Sec. \ref{sec:T2}.
Third, the cavity polarisation mode has a sub-optimal projection onto the atomic dipole moment, leading to a reduced value for $C$ (Sec. \ref{sec:supA}).
Considering these imperfections, the maximum photon-collection efficiency predicted by the analytical model [Eq. \eqref{eq:PS}] for the experiments presented in Secs. \ref{sec:exp_res} and \ref{sec:mult_phot} is ${P}^{exp.max}_{S} = 0.73(3)$ (Sec. \ref{sec:supA}).
While we do not measure  $P_S^{pure}$ in the experiment, we calculate the fraction of photons in the cavity output mode generated without prior spontaneous decay of the emitter to be $P_S^{pure}/P_S^{exp.max} = 0.53$.
Experimentally we achieve ${P}^{exp}_{S}=0.72(3)$ (Sec. \ref{sec:mult_phot}), saturating the analytic maximum to within uncertainty. 
Significantly higher photon collection efficiencies could only be achieved by changing properties of our cavity, by reducing either the losses or the waist. 

\section{Experimental details} \label{sec:exp_det}

\begin{figure}[t!]
	\vspace{0mm}
	\begin{center}
		\includegraphics[width=\columnwidth]{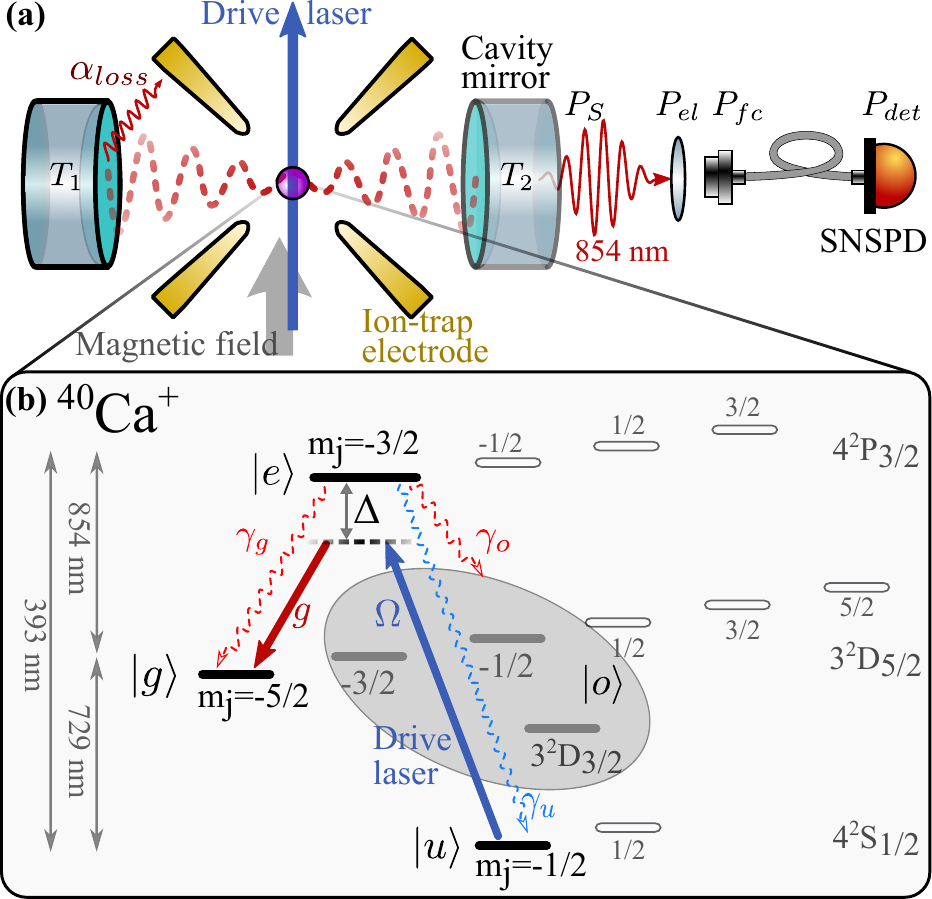}
		\vspace{0mm}
		\caption{
			\textbf{Experimental setup: photon generation from a trapped ion in a cavity.} 
			\textbf{(a)}  
			A $^{40}\text{Ca}^+$ ion in a linear Paul trap (shown electrodes are for radial, in plane, confinement) and at the focus, and a vacuum anti-node, of an optical cavity. The vacuum Rabi frequency of the ion cavity coupling is $g$. 
			A drive laser pulse with Rabi frequency $\Omega$ causes emission of an 854~nm photon into a vacuum mode of the cavity. 			
			$T_1$ and $T_2$ are mirror transmissions, 
			$\alpha_{loss}$ is the cavity round-trip loss, 
			$P_S$ is the probability for obtaining a photon in the cavity output mode,
			$P_{el}$ is the transmission of optical elements,
			$P_{fc}$ is the fibre coupling efficiency,
			$P_{det}$ is the detector efficiency.			
			\textbf{(b)}
			Atomic level scheme showing the CMRT for photon generation. 
			The cavity (red solid arrow) and drive laser (blue solid arrow) have a common detuning of $\Delta/2\pi=-403(5)$ MHz from the excited state. 
			Ground states other than $\ket g$ and $\ket u$ to which the excited state $\ket e$ can decay are grouped in state $\ket o$ (contained within grey oval).
			The spontaneous decay rates are $(\gamma_g,\gamma_u, \gamma_o)/2\pi = (0.45,10.74,0.30)$ MHz, with $\gamma_g + \gamma_u + \gamma_o = \gamma = 2\pi 11.49$ MHz.
		}
		\label{fig:exp}
		\vspace{-6mm}
	\end{center}
\end{figure}

Our emitter is a single $^{40}\text{Ca}^+$ ion in the centre of a linear Paul trap and at the position of the waist of an optical cavity that enhances photon emission on the 854 nm electronic dipole transition (Fig. \ref{fig:exp}). 
The 854 nm photon wavelength is well suited for a flying qubit, as optical fibre absorption rates (3 dB/km) allow for a few kilometers of travel.
Furthermore, the 854~nm transition in $^{40}\text{Ca}^+$ is uniquely suited amongst trapped-ion species for direct, efficient and low added-noise frequency conversion to the telecom wavelengths (0.2 dB/km fibre absorption rate) \cite{Walker2018, Bock2018, Krutyanskiy2019} and has allowed for the distribution of ion-photon entanglement over 50 km of optical fibre \cite{Krutyanskiy2019}.
\textcolor{black}{A summary of key technical advances made over the work of Ref. \cite{Krutyanskiy2019} is presented in Sec. \ref{sec:upgrades}.
}

Our 19.906(3) mm long, near-concentric cavity achieves a microscopic \textcolor{black}{waist and} macroscopic ion-mirror separation (10~mm), rendering the effect of mirror surface charges on the trapping potential negligible.
The mirror transmission values at 854~nm are $T_1=2.9(4)$ ppm and $T_2=90(4)$ ppm.
Other cavity photon loss is measured to be $L=23(4)$ ppm, of which $10(4)$ ppm is due to absorption and scattering in the mirror coatings and substrates and the remaining loss is attributed to imperfect cavity alignment.
The round-trip probability of unwanted photon loss in the cavity is $\alpha_{loss}=L+T_1=26(4)$ ppm. 
The cavity waist is centred on the ion via stick-slip piezo translation stages (Attocube) in three dimensions.
The ion is placed at an anti-node of the cavity vacuum field by tuning of the cavity position along the cavity axis. A small ($<10$ V) DC voltage on the corresponding translation stage, allowing for nm-scale displacements, is adjusted until the photon detection rate is maximised.
For details on the cavity parameters and how the relative position of ion and waist was determined, see Sec. \ref{sec:cavpara}.

Our photon generation scheme is based on a cavity-mediated Raman transition (CMRT) \cite{Keller2004, Barros2009}. 
For photons not entangled with the ion, the CMRT couples the initial state $\ket u \ket 0 = \ket{4^{2} S_{J{=}1/2},m_j{=}-1/2} \ket 0$, where the second ket vector describes the cavity photon number, to the metastable final state $\ket g \ket 1 = \ket{4^{2} D_{J{=}5/2},m_j{=}-5/2} \ket 1$ via the intermediate excited state $\ket e \ket 0 = \ket{4^{2} P_{J{=}3/2},m_j{=}3/2} \ket 0$.
The three atomic states are shown as thick black lines in Fig. \ref{fig:exp}b. 
States $\ket e$ and $\ket g$ have lifetimes of 6.9~ns and 1.1~s, respectively. 
Ground states other than $\ket g$ and $\ket u$ to which the excited state $\ket e$ can decay are grouped in a fourth state $\ket o$ (contained within grey oval in Fig. \ref{fig:exp}b).

A circularly polarised drive laser (100 Hz linewidth) addresses the $\ket u \leftrightarrow \ket e$ dipole transition at 393 nm with a detuning $\Delta/2\pi=- 403(5)$ MHz and a Rabi frequency $\Omega$. 
The cavity is detuned by the same amount from the 854 nm $\ket g \leftrightarrow \ket e$ dipole transition, and its length is actively stabilised via the Pound-Drever-Hall method to a laser at 806~nm wavelength.
Both this laser and the drive laser are stabilised to a common in-vacuum reference cavity; 
\textcolor{black}{see Sec. \ref{sec:cavpara} for details.} 
The ion-cavity coupling strength on the $\ket g \leftrightarrow \ket e$ transition is $g/2\pi = 1.25(1)\times \zeta \text{ MHz}$, where the geometric factor $\zeta \le 1$ is the projection of the cavity polarisation onto the atomic dipole moment. 
A 4.23~G magnetic field, set by rings of permanent magnets, is aligned perpendicular to the cavity axis and parallel to the propagation direction of the drive field.
The atomic quantisation axis is chosen to be parallel to the magnetic field axis, and photon polarisation along this axis is denoted horizontal ($H$). 
The polarisation of the photon generated in the cavity is vertical ($V$), as determined by the projection of the $\ket g \leftrightarrow \ket e$ dipole moment onto the plane perpendicular to the cavity axis, corresponding to $\zeta=\sqrt{0.5}$ and $g_{exp}/2\pi = 0.88(1)$~MHz.
The effective coupling strength of the CMRT is $\Omega_{e\!f\!f}= g \Omega/(2\Delta)$.
The CMRT competes with spontaneous decay with an effective rate $\gamma_{e\!f\!f} =[\Omega/(2\Delta)]^2 \gamma$, where $\gamma = 11.49(3)$ MHz is the decay rate of $\ket e$ \cite{Jin1993}.
For $\Omega/2\pi = 14$ MHz, the lowest drive strength used in the experiment of Sec. \ref{sec:exp_res}, the effective drive strength, cavity decay rate and effective atomic decay rate are given by $(\Omega_{e\!f\!f}, \kappa, \gamma_{e\!f\!f}) = 2\pi \times (16.2, 70, 3.9)$  kHz. 

Each photon generation attempt consists of the following sequence:
1) Doppler cooling and optical pumping to $\ket u$;
2) sideband-resolved laser cooling at 729~nm of the ion axial mode ($\omega_z/2\pi = 0.92$ MHz) and two radial modes ($\omega_{r1}/2\pi = 2.40$ MHz and $\omega_{r2}/2\pi = 2.44$ MHz) yielding mean phonon numbers below one in each mode; 
3) a drive laser pulse, ideally triggering the emission of a photon by the ion into the cavity;
4) photon detection with a superconducting nanowire single-photon detector (SNSPD) and time-tagging of detection events. 
In addition, for the experiment of Sec. \ref{sec:ent}, we carry out an additional step of detection of the ion qubit and photon polarisation state (involving a second SNSPD) for characterisation of the ion-photon entangled state, as detailed in that section.
The complete experimental sequence for generating a single photon lasts 8 ms, dominated by 7 ms of sideband cooling.
A detailed laser pulse sequence can be found in Sec. \ref{sec:sequence}.

Detected photons are the ones that exit via the right-hand mirror in Fig. \ref{fig:exp}, pass some passive optical elements with probability $P_{el} = 0.97(1)$, are coupled into a few-meter-long single-mode optical fiber with an efficiency $P_{fc} = 0.81(3)$ and are finally detected with an SNSPD that has an efficiency $P_{det} = 0.87(2)$ and free-running dark counts 0.3(1) per second. 
\textcolor{black}{
	The dark count rates are negligible relative to background light rates (5-20 counts/second)
	and data rates (peak 60 kcounts/second).}
The path efficiency is therefore $P_{path} = P_{el}P_{fc}P_{det} = 0.68(3)$. 
The total detected photon probability is $P_{tot}=P_S P_{path}$ where $P_S$ is the probability for obtaining a photon in the cavity output mode. 

For a quantitative prediction of $P_S$, for a given set of parameters of the CMRT, we numerically solve a model based on a master equation comprising 18 electronic states in \Ca and two frequency-degenerate  modes of the cavity \cite{Brandstatter2013}.
All parameters in the model are independently measured or estimated via calibration experiments such that there are no free parameters within calibration uncertainties. 
The numerical model includes a broadening of the \textcolor{black}{CMRT} linewidth by 10 kHz due to imperfect length stabilisation; \textcolor{black}{ see Sec. \ref{sec:cavpara} for details.}

\section{Single-photon results}
\label{sec:exp_res}

\begin{figure}[t]
	\vspace{0mm}
	\begin{center}
		\includegraphics[width=\columnwidth]{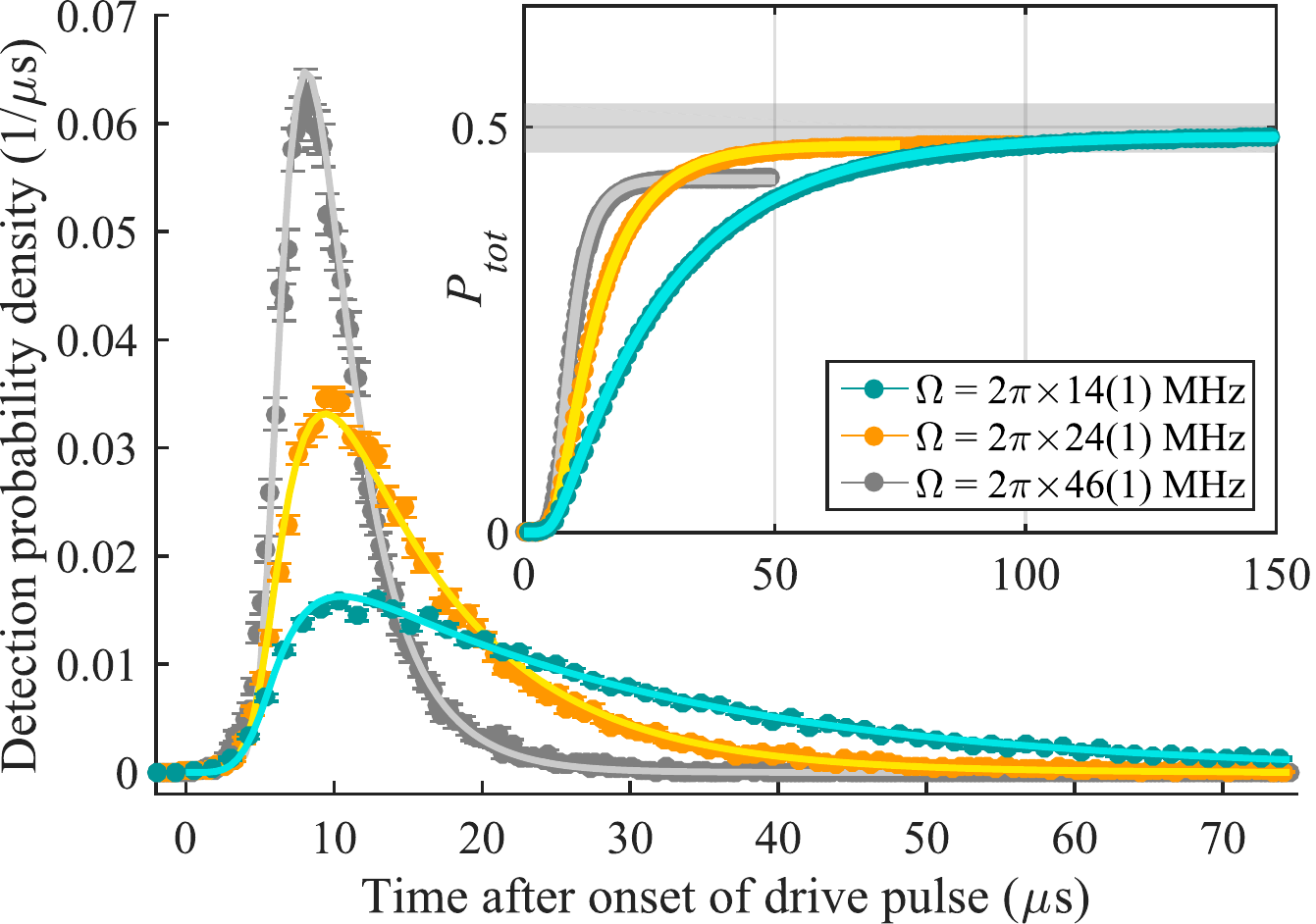}
		\vspace{0mm}
		\caption{
			\textbf{Single-photon wavepackets and efficiency.} 
			Histograms of photon arrival times for different Rabi frequencies $\Omega/2\pi = [14(1),24(1),46(1)]$ MHz of the drive laser, normalised by the number of attempts $k = 50000$ and by the time bin width $\delta_t = (1.2, 0.6, 0.3)~\upmu$s.
			\textbf{Inset}: Integrated wavepackets, yielding efficiencies (averaged over a measurement time of about 8 min) $P_{tot} = [0.490(3), 0.478(3), 0.437(3)]$.
			All plots: dots are experimental data, lines correspond to numerical simulations. Error bars \textcolor{black}{correspond to one standard deviation} based on Poissonian photon counting statistics.
			Grey shaded area: $P^{exp.max}_S \times P_{path}$, i.e., maximum achievable total efficiency from the analytic model [calculated via Eq. \eqref{eq:PS}] given our detection-path efficiency.
		}
		\label{fig:WP}
		\vspace{-6mm}
	\end{center}
\end{figure}

In the first experiment, the total detected photon probability $P_{tot}$ is measured for three drive-laser Rabi frequencies $\Omega/2\pi = [14(1),24(1),46(1)]$ MHz, and the time-tags of photon detection events are recorded. 
The temporal profiles of the detected single photons are shown in Fig. \ref{fig:WP}. 
These single-photon wavepackets are presented as a probability density $p_{d}(t) = N_d/(k\cdot \delta_t)$, where $N_{d}$ is the number of detection events registered in a time bin $\delta_t$ and $k$ is the number of attempts to generate a photon. 
The total detected probability $P_{tot} = P_S P_{path}$ is calculated by integrating the wavepacket and plotted in the inset of Fig. \ref{fig:WP}.

For $\Omega/2\pi = 14$ MHz, the lowest drive strength, the highest detected probability is obtained: $P_{tot}=0.490(3)$.
Out of the 50,000 photon generation attempts, 24,358 cases led to exactly one photon being detected in the time interval of the drive laser pulse.
In 28 cases a second photon was detected, consistent with the background light level of 5 per second. 
After factoring out the independently measured path efficiency $P_{path} = 0.68(3)$, this detected single-photon probability corresponds to a collection probability of $P^{exp}_S=0.72(3)$ out of the cavity.
The measured collection probability $P^{exp}_S$ is consistent with the prediction of the numerical model $P^{num}_S=0.72(3)$ and with the maximum analytical value $P^{exp.max}_S = 0.73(3)$, where the error in both predicted values is due to the uncertainty in the measured values of $T_2$ and $\alpha_{loss}$.
We thus conclude that, to within the calibration uncertainty of the detection path, we reach the maximum collection probability in our system given only by atomic quantities and the cavity parameters of transmission, loss, waist and direction.

Significant improvements in $P_S$ would have to involve changing the cavity properties and could not be achieved by, e.g., further reducing the spatial localisation of the ion, modifying the temporal or spectral properties of the drive laser or choosing a different drive scheme. Numerical simulations show that improving the cavity length stability could at most increase $P_S$ by $0.01$.  
Further discussion on ways to improve the efficiency by changing the setup is postponed until Sec. \ref{sec:improvements}.	

\textcolor{black}{
Numerical simulations show that decreasing $\Omega/2\pi$ from our lowest chosen value (14 MHz) by a factor of 1/3 would increase the achievable collection probability by at most 0.3\%. 
As this increase is below the statistical resolution of our presented measurements, no systematic studies have been performed in that regime. 
In general, the collection probability saturates as $\Omega$ tends to 0, and any slight improvements come at the cost of increasingly high stability requirements for the laser frequency and cavity lock.}
As $\Omega$ is increased, the photon collection probability $P_S$ is reduced because of the increased probability for spontaneous scattering: the effective spontaneous decay rate $\gamma_{e\!f\!f}$ increases faster with $\Omega$ than the coherent coupling rate $\Omega_{e\!f\!f}$, leading to a higher probability of the state ending in $\ket g \ket 0$ or $\ket o \ket 0$.
While increasing $\Omega$ decreases $P_S$, a stronger drive produces photons that are temporally shorter, which could be advantageous for experiments demanding higher photon rates. 
\textcolor{black}{Schemes to shape and chirp the drive-laser pulse \cite{Nisbet-Jones2011, Morin2019} could be used to minimise the reduction in $P_S$ that occurs for faster driving (larger $\Omega$).
}

\section{Ion-photon entanglement}
\label{sec:ent}

\begin{figure}[t]
	\vspace{0mm}
	\begin{center}
		\includegraphics[width=\columnwidth]{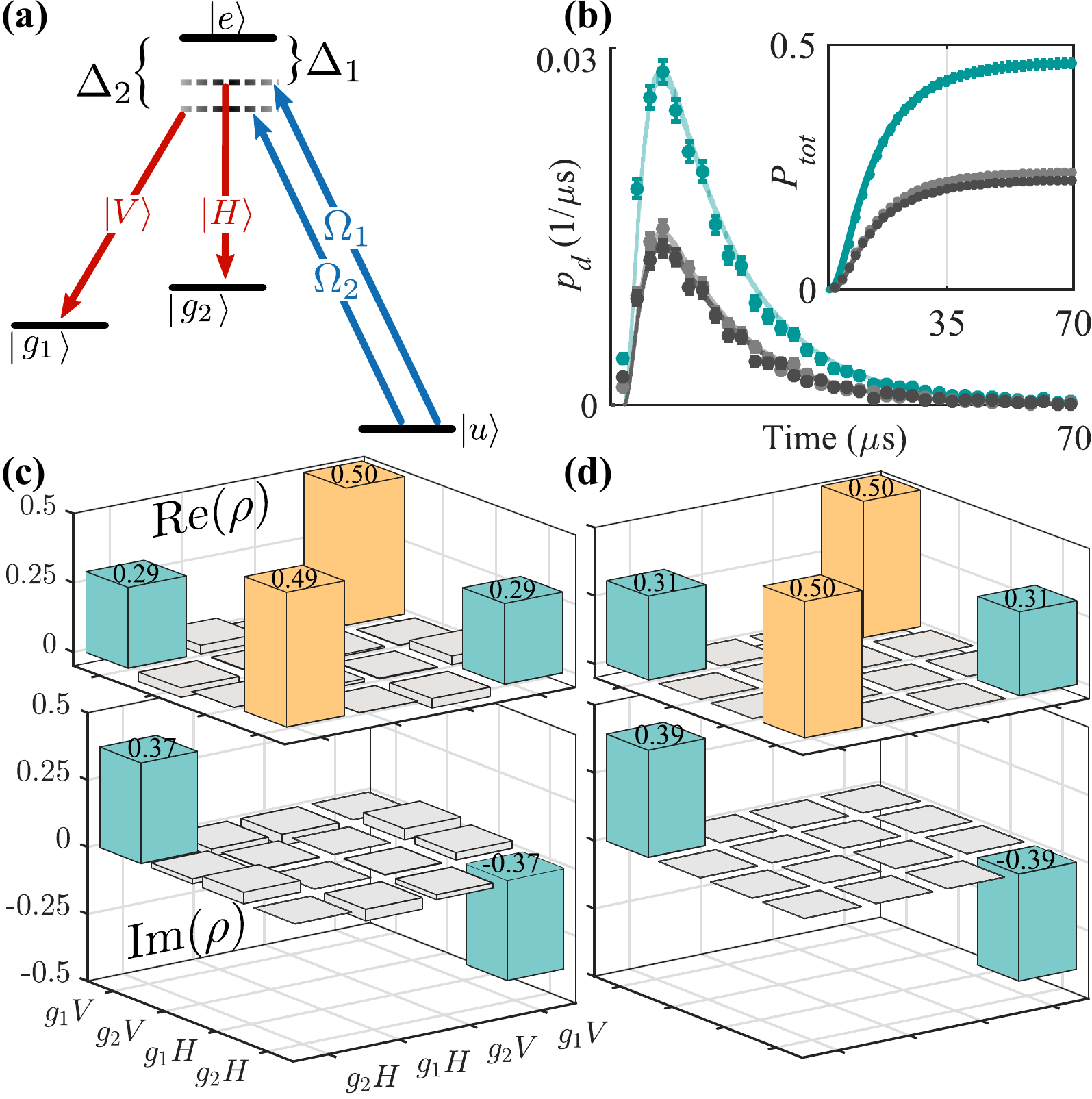}
		\vspace{0mm}
		\caption{
			\textbf{Generation and characterisation of the ion–photon entangled state.} 
			\textbf{(a)} Atomic level scheme showing the bichromatic CMRT that generates the ion-photon entangled state $\ket{\Psi(\theta)}=\frac 1{\sqrt 2}(  \ket {g_1} \ket V + e^{i\theta} \ket{g_2} \ket H)$.
			The frequency difference of the detunings $\Delta_1$ and $\Delta_2$ is equal to the Zeeman splitting between $\ket{g_1} = \ket{D_{J=5/2},m_j=-5/2}$ and $\ket{g_2} = \ket{D_{J=5/2},m_j=-3/2}$.
			\textbf{(b)} Wavepackets in the $H/V$ (grey/black) polarisation basis and their sum (green). The photon counts of all three ion-basis measurements were added up.
			Dots with error bars: measured data; lines: numerical simulations.
			Error bars represent one standard deviation, based on Poissonian photon statistics.
			\textbf{Inset:} integrated probabilities, yielding 0.224(4) for $H$ and 0.239(4) for $V$.
			\textbf{(c) Data:} Real (top) and imaginary (bottom) part of the tomographically reconstructed density matrix $\rho$.
			\textbf{(d) Theory:} Real (top) and imaginary (bottom) part of the density matrix of $\ket{\Psi(\theta=0.91)}$.
		}
		\label{fig:entang}
		\vspace{-6mm}
	\end{center}
\end{figure}

Entanglement between the photon and ion is generated via a bichromatic CMRT that was developed and first demonstrated in Ref. \cite{Stute2012}. 
Here, a second drive-laser field at a different frequency is added, detuned from the frequency of the original drive-laser field by the Zeeman splitting ($Z/2\pi=7.1$~MHz) between $\ket{g_1} = \ket g = \ket{D_{J=5/2},m_j=-5/2}$ and neighbouring state $\ket{g_2} = \ket{D_{J=5/2},m_j=-3/2}$; see Fig. \ref{fig:entang}a. 
Consequently, a second Raman process is simultaneously driven that generates a horizontally polarised ($H$) cavity photon. 
In the case where the two processes occur with equal probability, the initial state $\ket u \ket 0$ is ideally transferred to the final, maximally entangled state $\ket{\Psi(\theta)}=\frac 1{\sqrt 2}(  \ket{g_1} \ket V + e^{i\theta} \ket{g_2} \ket H)$, where the phase $\theta$ is set by the relative phase of the two laser fields and can be tuned \cite{Stute2012}.
A qubit encoded in the photon's polarisation is therefore entangled with a qubit encoded in the ion's electronic states $\ket{g_1}$ and $\ket{g_2}$.
The maximum probabilities for collecting photons via each Raman process, when generated separately in our system, calculated using Eq. \eqref{eq:PS}, are $P^V_S=0.728(30)$ and $P^H_S=0.717(30)$, for the $V$ and $H$ photon, respectively (Sec. \ref{sec:PSexp}). 
While we do not have an analytic expression for the maximum probability for generating an ion-entangled photon, numerical simulations suggest that the value tends to the mean of $P^V_S$ and $P^H_S$ as the total Rabi frequency tends to 0.  

The individual Rabi frequencies $\Omega_1$ and $\Omega_2$ are set by varying the powers of the two drive-laser fields until the probability for detecting photons is balanced in the $H/V$ polarisation basis, to within statistical uncertainty.
The total Rabi frequency $\Omega=\sqrt{\Omega_1^2+\Omega_2^2}=2\pi\times22(1)$~MHz is then determined by measuring the AC-Stark shift induced on the 729~nm transition by the combined CMRT fields; for details on the calibration, see Sec. \ref{sec:calibration}.

For the characterisation of the ion-photon state, the first step is to move the $\ket{g_1}$ electron population to the state $\ket u$ via a $\pi$-pulse using a laser at 729 nm.  
That is, the D-manifold qubit, consisting of states $\ket{g_1}$ and $\ket{g_2}$, is mapped onto an S-D `optical' qubit, consisting of states $\ket u$ and $\ket{g_2}$, over which full quantum control is well established \cite{Schindler2013, Friis2018}. 
The second step is to perform full quantum-state tomography to reconstruct the two-qubit ion-photon state as in Ref. \cite{Stute2012}; see Sec. \ref{sec:state_characterisation} for experimental details on the laser pulse sequence and state reconstruction.

For quantum state tomography, 45,000 attempts were made to generate an ion-photon entangled state over a ten minute experimental run. 
In 20,788 cases at least one photon detection event occurred during the drive laser pulse. 
In the four cases where more than one detection event occurred during a drive laser pulse, the event that occurred first was taken as the result.
The measured detection probability of the ion-photon entangled state is therefore $P_{tot} = 0.462(3)$, which, given the detection-path efficiency, corresponds to a collection probability of $P_S = 0.69(3)$ at the cavity output.
This measured value of $P_S$ is consistent with the value $0.70(3)$ predicted by master equation simulations for a bichromatic drive field with Rabi frequencies $\Omega_1/2\pi = 14.2$ MHz and $\Omega_2/2\pi = 16.8$ MHz. These values for $\Omega_{1,2}$ are consistent with the measured value of $\Omega$ and predict photon wavepackets that closely match the ones observed in experiment. 
Fig. \ref{fig:entang}b presents photon wavepackets for the cases where $H$ and $V$ photons were detected (ignoring the detected  state of the ion qubit), yielding total integrated probabilities of $P_H=0.239(4)$ and $P_V=0.224(4)$, respectively. The imbalance of these probabilities is beyond the statistical resolution of the aforementioned calibration step. 

The tomographically reconstructed density matrix of the ion-photon state $\rho$ (Fig. \ref{fig:entang}c) has a fidelity of $F=\text{Tr}(\ket{\Psi}\langle\Psi|\rho)=0.966(5)$ with the maximally entangled state $\ket{\Psi(\theta=0.91})$ and a purity $\mathcal{P}=\text{Tr}(\rho^2)=0.948(9)$.  
The statistical uncertainties here are obtained via the Monte-Carlo method \cite{Efron1986} based on photon counting statistics.
The measured  state fidelity is not significantly limited by the imbalanced polarisation components nor by background photon counts. 
Specifically, the (unitary) simulations of Fig. \ref{fig:entang}b predict a fidelity with respect to a maximally entangled state of 0.9992, while background counts at the measured rate of 20 counts per second further limit the maximum achievable fidelity to 0.9974(7) (Sec. \ref{sec:state_characterisation}).
It is straightforward to show that the maximum fidelity of an arbitrary state  $\rho_{arb}$ with any pure state is given by the square root of the purity of $\rho_{arb}$, that is, $F\leq \sqrt{P}$. 
Our reconstructed state saturates the aforementioned bound up to a difference $\sqrt{P}-F=0.008(7)$, and therefore our total state infidelity  [$1-F=0.034(5)]$ is almost entirely due to the lack of purity. 
Possible causes of the lack of purity in the experimental state are the cumulative effects of imperfections in the 729~nm laser pulses used in the ion qubit state analysis as well as imperfections in the polarisation analysis of the photon.  
A detailed error budget at the percent level is beyond the scope of this work, requiring, e.g., improved measurement statistics for verification.  

\section{Multi-photon states}
\label{sec:mult_phot}

A promising application of a near-deterministic interface between photons and quantum emitters  is to generate the entangled multi-photon states that enable scalable quantum computing \cite{Lindner2009, Economou2010, Schwartz2016}, as recently demonstrated for up to five sequential photons from a quantum dot \cite{Schwartz2016}, or long distance quantum networking \cite{Munro2015, Azuma2015, Borregaard2020}. 
While the trapped ion platform has great potential to enable the aforementioned applications, the generation and detection of photon trains so far been limited by low single-photon detection efficiencies, in comparison to those achieved in the present work.

We report now on an experiment in which each experimental sequence consists of one instance of sideband cooling followed by 15 attempts to generate a vertically polarised photon, without further cooling. The 15 attempts (15 Raman laser pulses, each lasting $200~\upmu$s) are spaced by only 60 $\upmu$s, during which time only electronic state reinitialisation, via repumping and optical pumping, is performed. 
By limiting the sideband cooling time in this way, the total laser pulse sequence for attempting to generate a train of 15 photons is reduced by an order of magnitude (from 125\,ms to 12.5\,ms). 

\begin{figure}[t]
	\vspace{0mm}
	\begin{center}
		\includegraphics[width=\columnwidth]{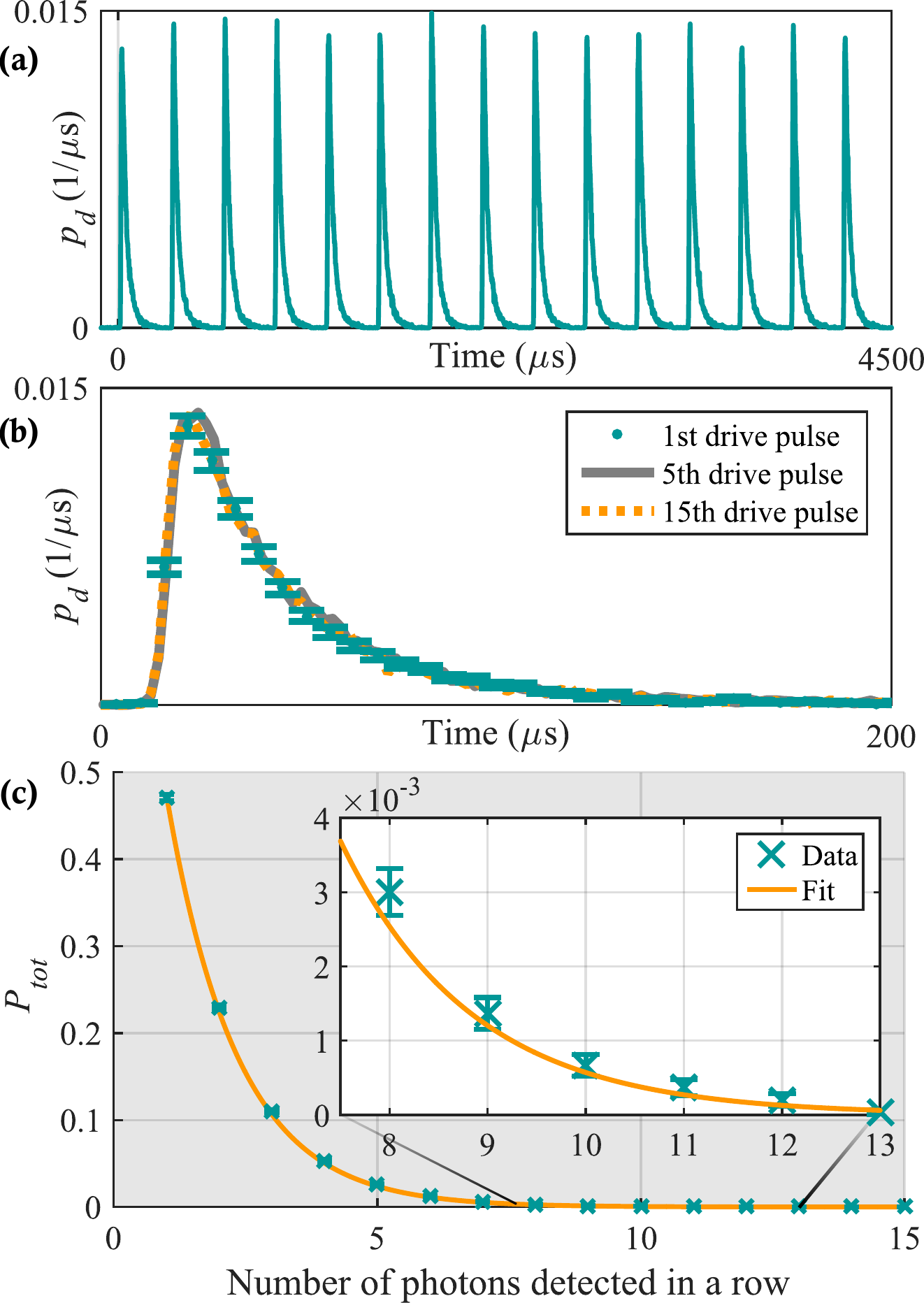}
		\vspace{0mm}
		\caption{
			\textbf{Photon trains: generation and detection of sequential photons} 
			\textbf{(a)} Measured wavepackets of 15 consecutive photon generation attempts for the experimental configuration of Fig. \ref{fig:exp} and $\Omega/2\pi\approx 12$ MHz. 
			\textbf{(b)} Temporal overlap of the 1st, 5th, and 15th wavepacket in (a).
			\textbf{(c)} Probability of detecting a photon in $n$ consecutive wavepackets, starting from the first wavepacket, for a single-photon probability of 0.47. Orange line: weighted fit of the form $p^x$, with $p=0.474(2)$ the best fit value. 
		}
		\label{fig:multphot}
		\vspace{-6mm}
	\end{center}
\end{figure}

A total of 30,000 attempts were made to generate trains of 15 photons over a twelve-minute experiment \cite{duration}. 
The 15 detected single-photon wavepackets are presented in Fig. \ref{fig:multphot}a and reveal no statistically significant differences across the train, as shown in Fig. \ref{fig:multphot}b.
When averaged over all 15 wavepackets, the detected single photon probability is $P_{tot}=0.471(4)$. 
The measured probabilities for detecting $n$ sequential photons, beginning with the one in the first time window in the train, are presented in Fig. \ref{fig:multphot}c and are fit with an exponentially decaying sequential photon probability of $[0.474(2)]^n$. 
Over the experiment we observed 90 instances of single photons in each of the first 8 windows, 20 instances of single photons in the first 10 windows, and 1 instance of all 15 time windows being full. 
Significant further improvements in the multi-photon detection rates in our system should be possible by increasing the attempt rate, e.g., by using periodic fast ground-state cooling techniques (requiring a few tens of microseconds) \cite{Lechner2016, Joshi2020}.

Ultimately, the limit to the rate at which single photons can be collected from our system is set by the decay rate $\kappa/2\pi = 70$ kHz of our two-centimeter-long cavity. 
Recall that, for a given quantum emitter, the cavity waist and unwanted photon loss $\alpha_{loss}$ determine the required optimum output mirror transmission $T_2^{opt}$ for maximal photon collection probability. 
After that optimisation, the cavity length $l$ remains a free parameter. 
Since $\kappa$ is proportional to $1/l$, the shorter the cavity, the higher the value of $\kappa$, and therefore the higher the potential rate of single-photon production. 
Integrating trapped ions with fiber cavities of sub-mm length is a promising path to increase the single-photon generation rate, and significant progress has been recently made in that direction \cite{Takahashi2020, Kobel2021}.

The first step towards generating entangled photon trains in an ion-cavity system would be to perform coherent manipulation of the ion qubit after photon generation, instead of the incoherent reinitialisation performed here. 
\textcolor{black}{That step should be straightforward to achieve using the existing tools (729~nm laser) for manipulating the ion qubit available in the present work.  
The second step would be to reduce spontaneous scattering (reexcitation) events during the photon generation process, which destroy entanglement between sequentially generated cavity photons from a single ion. Indeed, in our current experimental configuration, these spontaneous scattering events would already strongly limit the probability with which two sequentially entangled photons could be generated. 
Those same spontaneous scattering events also limit the extent to which the emitted photons are indistinguishable, as we \cite{Meraner2020} and others \cite{Walker2020} have recently studied in an ion-cavity system.
Sec. \ref{sec:improvements2} presents schemes to increase the fraction of emitted cavity photons without prior spontaneous emission in our existing system to over 0.8, which we expect to allow for the observation of entanglement between several sequential photons. 
Detailed numerical simulations are required to determine the exact number of sequentially entangled photons that could be observed in our current system, for a given detection probability, and we leave this for future work.
Sec. \ref{sec:improvements} present schemes to further reduce the effects of spontaneous scattering in future experimental systems. 
}

\section{Future systems}
\label{sec:improvements}

We have saturated the maximum probability for obtaining a photon in the output mode of the optical cavity $P_S^{exp.max}$, up to the statistical uncertainty in our data of a few percent (Sec. \ref{sec:exp_res}).
Consequently, significant improvements can only be achieved by changing properties of the cavity ($w_0$, $\alpha_{loss}$) or quantum emitter.
We now present three parameter combinations for future experimental systems that would enable a significant improvement: specifically, over 0.9 for both $P_S^{bound}$ and $P_S^{pure} / P_S^{bound}$, where the latter is the fraction of photons occurring without reexcitation events (Sec. \ref{sec:model}). 
As a basis for the calculations we take the parameters of $w_0$, $\alpha_{loss}$ and emitter structure from our present system, assuming a polarisation projection $\zeta = 1$, unless otherwise indicated.
The first parameter combination consists of a ten times lower value of 2.7~ppm for $\alpha_{loss}$ combined with a value of 25~ppm for $T_2$. 
Such a low $\alpha_{loss}$ was reported in Ref. \cite{Rempe1992}. 
The second parameter combination consists of a three times smaller value of $3.9~\upmu$m for $w_0$ combined with a value of 252~ppm for $T_2$. 
Macroscopic cavities with waists down to 2.44 $\upmu$m have been realised \cite{Nguyen2018}, albeit with far higher losses than ours. 
The third parameter combination consists of ten ions maximally coupled to the cavity in superradiant entangled states, and a value of 252~ppm for $T_2$.
Superradiant states of two emitters coupled to a cavity have been demonstrated \cite{Casabone2015} and systems where the emitter string is parallel to the cavity axis \cite{Cetina2013, Begley2016} offer a path to coupling many entangled single emitters to a single cavity mode.
Indeed, it should be feasible to maximally couple up to three ions in superradiant states to our existing cavity, allowing for a significant improvement in $P_S^{pure} / P_S^{bound}$, as detailed in Sec. \ref{sec:improvements2}.

\section{Conclusion}
\label{sec:conclusion}

This work presented a significant advance in the collection efficiency $P_S$ of photons entangled with an ion.  
This advance was made possible by a cavity with a microscopic waist, total unwanted photon loss of 26(4)~ppm and an outcoupling mirror transmission that achieves close to the optimal compromise between photon emission into the cavity and escape into the output mode. 
These are the key parameters which, together with the level structure of the quantum emitter, determine the optimal design for emitter-cavity systems with maximum photon collection efficiency.  
We have shown that our efficiency saturates, to within a few percent, the maximum value allowed by our cavity and emitter parameters. 
We thus could not achieve significantly higher efficiencies by, e.g., further optimising the photon generation scheme, changing to another scheme or further reducing the ion's spatial extent. 
Significant improvements would require a different value for the output coupling mirror transmission combined with either a smaller waist, lower unwanted losses or cavity-coupled superradiant states of multiple ions, or some combination of all these approaches. 

An interesting future prospect is to use the interface presented here to access new regimes in the measurement of trapped-ion qubits, which is currently done via the electron shelving technique, during which large numbers of photons are scattered. Using the near-deterministic photon collection presented here, it may be possible to reliably determine the ion state whilst only scattering a few photons and thereby minimally disrupting the ion's motional state. Besides reducing the need for re-cooling, motionally non-destructive ion-qubit state readout could be feasible, opening new possibilities for the storage and engineering of quantum information in motional degrees of freedom.

\begin{acknowledgments}

We thank Rainer Blatt for helpful discussions and for providing the laboratory space; Stefan Haslwanter for designing and fabricating our ion-trap system; Tiffany Brydges and Christine Maier for their support with a shared laser system; Takao Aoki for helpful discussions.

This work was financially supported by the START prize of the Austrian FWF project Y 849-N20, the Army Research Laboratory Center for Distributed Quantum Information via the project SciNet under Cooperative Agreement Number W911NF-15-2-0060, the Institute for Quantum Optics and Quantum Information (IQOQI) of the Austrian Academy Of Sciences (OEAW) and the European Union’s Horizon 2020 research and innovation program under grant agreement No 820445 and project name ‘Quantum Internet Alliance’. The European Commission is not responsible for any use that may be made of the information this paper contains.
Ben Lanyon is a Fellow in the CIFAR Quantum Information Science Program.
\end{acknowledgments}

\section*{Author Contributions}

All authors contributed to the design, development and characterisation of the experimental systems. 
In particular, J.S. focused on the optical cavity and general design of the experimental system,
V.Krc. on the ion trap, 
M.M. on the experiment control, 
V.Kru. and B.P.L. on all aspects. 
Experimental data taking was done by J.S., V.Krc. and V.Kru. 
Data analysis and interpretation was done by J.S., B.P.L. and V.Kru.
T.E.N. developed the code for the numerical simulations.
The manuscript was written by B.P.L. and J.S.
All authors commented on the manuscript. 
The project was conceived and supervised by B.P.L.

\appendix

\renewcommand{\appendixname}{APPENDIX}

\section{ANALYTIC MODEL FOR THE LIMIT TO $P_S$}\label{sec:supA}
This section gives details on the model developed in Ref. \cite{Goto2019} which is used in this paper to calculate the maximum photon collection probability $P_S$ [given by Eq. \eqref{eq:PS}] in our experiment and to identify paths for improvements.
In Sec. \ref{sec:defs}, the parameters used in the model are defined.
In Sec. \ref{sec:PSexp}, the the maximum photon collection probability in our experiment $P^{exp.max}_{S}$ is explicitly calculated.
In Sec. \ref{sec:PSopt}, the specific form of Eq. \eqref{eq:PSopt} (given in Sec. \ref{sec:model}) is explained.

\subsection{Definitions of parameters} 
\label{sec:defs}

The atom-cavity coupling strength is given by
\begin{equation}
	g = \sqrt{\frac{c\gamma_{g}}{2l\widetilde{A}_{e\!f\!f}}}\zeta
	\label{eq:g}
\end{equation}
where $c$ is the vacuum speed of light, $\gamma_g$ the spontaneous atomic decay rate of the transition coupled by the cavity, $l$ the cavity length and $\widetilde{A}_{e\!f\!f} = \frac{A_{e\!f\!f}}{\sigma}$ is the cavity mode area ${A}_{e\!f\!f}=w_0^2 \pi /4$ at the point of the atom, normalised by the atomic absorption cross section $\sigma = 3\lambda^2/2\pi$, with $w_0$ the cavity waist and $\lambda$ the wavelength of the cavity-coupled transition.
The projection of the cavity polarisation onto the atomic dipole moment is accounted for by $\zeta \le 1$.
If multiple ($N$) atoms are coupled to the cavity in a super-radiant entangled state, $g$ increases by a factor $\sqrt N$ \cite{Lamata2011}.

The cavity decay rate is given by
\begin{equation}
	\kappa=\kappa_{in}+\kappa_{ext}=\frac{c}{4l}({\alpha_{loss}+T_2})
	\label{eq:kappa}
\end{equation}
where the external decay rate is $\kappa_{ext} = \frac{c}{4l}T_2$ and corresponds to wanted decay into the output mode, so transmission through the cavity output mirror that has a transmission $T_2$; see Fig. \ref{fig:model}.
The internal decay rate is $\kappa_{in} = \frac{c}{4l}\alpha_{loss}$, where $\alpha_{loss}$ is the unwanted cavity loss, containing all photon loss mechanisms in the cavity except for $T_2$.

The cooperativity is given by
\begin{equation}
	C = \frac{g^2}{2\kappa\gamma} = \frac{1}{\widetilde{A}_{e\!f\!f}(\alpha_{loss}+T_2)}\frac{\gamma_g}{\gamma}\zeta^2
	\label{eq:coop2}
\end{equation}
with $\gamma = \gamma_g + \gamma_u + \gamma_o$ the spontaneous decay rate of the excited state.
Decay rates from $\ket e$ to ground states other than $\ket g$ or $\ket u$ are summed up in $\gamma_o$.
The definition of $C$ includes the projection factor $\zeta$ from the definition of $g$.

\subsection{Calculation of $P_S^{bound}$ for our experimental setup}
\label{sec:PSexp}

The maximum possible probability for collecting a photon from our ion-cavity system, for a given experimental configuration, is calculated via Eq. \eqref{eq:PS}.
For the experimental configuration shown in Fig. \ref{fig:exp}, the parameters necessary for the calculation of ${P}^{exp.max}_{S} = 0.73(3)$ are now provided.

\begin{figure}[t]
	\vspace{0mm}
	\begin{center}
		\vspace{0mm}
		\includegraphics[width=0.6\columnwidth]{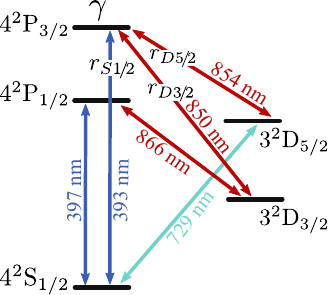}
		\caption{
			\textbf{\Ca level scheme with branching ratios and additional transitions used for cooling.} 
			Total decay of $P_{3/2}$ \cite{Jin1993}: $\gamma/2\pi = 11.49(3)$ MHz.
			Branching ratios \cite{Gerritsma2008}:
			$r_{D1\!/\!2} = 0.9347(3)$,
			$r_{D3\!/\!2} = 0.00661(4)$,
			$r_{D5\!/\!2} = 0.0587(2)$ .
		}
		\label{fig:full_level}
		\vspace{-6mm}
	\end{center}
\end{figure}

The atomic decay rates are calculated as follows:
\begin{center}
	$\gamma_{g} = \gamma\times {r_{D5\!/\!2}}\times (G_{-\frac5 2})^2$,\\
	$\gamma_u = \gamma\times {r_{S1\!/\!2}}$,\\
	$\gamma_{o} = \gamma\times r_{D5\!/\!2} [(G_{-\frac3 2})^2 + (G_{-\frac1 2})^2] + \gamma\times r_{D3\!/\!2}$,
\end{center}
yielding $(\gamma_g,\gamma_u, \gamma_o)/2\pi = (0.45,10.74,0.30)$ MHz. 
The decay rate $\gamma$ and the branching ratios $(r_{D5\!/\!2}, r_{D3\!/\!2},r_{D1\!/\!2})$ to the corresponding fine-state manifolds are given in Fig. \ref{fig:full_level}.
The Clebsch-Gordan coefficients $G_m$ for transitions $\ket{P_{J{=}3/2},m_j{=}-3/2} \rightarrow \ket{D_{J{=}5/2},m}$ with final Zeeman sub-level $\ket m$ are given by
$G_m = (\sqrt{10/15},-\sqrt{4/15},\sqrt{1/15})$, with $m=(-\frac5{2},-\frac3{2},-\frac1{2})$.

For the experiment presented in Sec. \ref{sec:exp_res}, the cavity couples the transition to $m=-\frac 5{2}$.
In free space this transition generates a $\sigma^+$-polarised photon which is projected to vertical ($V$) polarisation within the cavity, resulting in $\zeta = \sqrt{0.5}$.
With the cavity parameters given in Table \ref{table:cav}, at the wavelength $\lambda=854$ nm, the following values are obtained: 
$\widetilde{A}_{e\!f\!f} = 34.2$,
$g/2\pi = 0.88(1)$ MHz, 
$C = 0.49(1)$.
The corresponding probabilities are $P_{in}=0.940(2)$, $P_{esc} = T_2/(\alpha_{loss}+T_2) = 0.78(3)$, so Eq. \eqref{eq:PS} yields ${P}^{exp.max}_{S} = P_{in} P_{esc} = 0.728(30)$.

For the generation of ion-photon entanglement, presented in Sec. \ref{sec:ent}, the cavity additionally couples the transition to $\ket{D_{J=5/2},m=-3/2}$.
In free space this generates a $\pi$-polarised photon which is projected to horizontal ($H$) polarisation within the cavity, resulting in $\zeta$ = 1 .
For that transition the maximum photon collection probability is $0.717(30)$.

\subsection{Upper bound of $P_S$ for optimal mirror transmission $T_2$}
\label{sec:PSopt}

In Eq. \eqref{eq:PS} both $P_{in}$ and $P_{esc}$ depend on $T_2$. 
Maximising Eq. \eqref{eq:PS} with respect to $T_2$ yields Eq. \eqref{eq:PSopt} for the optimum output-mirror transmission
\begin{equation}
	T^{opt}_2 = \alpha_{loss} \sqrt{1+\beta\frac 1{\alpha_{loss}}\frac 2{\widetilde{A}_{e\!f\!f}}},
	\label{eq:T2opt}
\end{equation}
with $\beta = \frac{\gamma_g}{\gamma-\gamma_u}\zeta^2$.
Eq. \eqref{eq:PSopt} corresponds to Eq. (34) in Ref. \cite{Goto2019}, but the latter is expressed in terms of the internal cooperativity
$C_{in} = \frac{g^2}{2\kappa_{in}\gamma} =  \frac{1}{\widetilde{A}_{e\!f\!f}\alpha_{loss}}\frac{\gamma_g}{\gamma}\zeta^2$.
Eq. \eqref{eq:PSopt} follows directly from Eq. (34) in Ref. \cite{Goto2019} by substituting $2C_{in}/(1-r_u) = \beta\frac{2}{\widetilde{A}_{e\!f\!f}\alpha_{loss}}$. 
We introduced the parameter $\beta$ to simplify Eq. \eqref{eq:PSopt} by making it independent of the number of atomic levels and to capture the different cases discussed in Secs. \ref{sec:model} and \ref{sec:improvements}:
\begin{itemize}
	\item For a 3-level atom the total spontaneous decay rate of the excited state $\ket e$ is given by $\gamma = \gamma_g+\gamma_u$, so $\beta=1$ (for $\zeta=1$).
	\item For a quantum emitter such as ours (Fig. \ref{fig:exp}) the excited state can decay to levels other than $\ket g$ and $\ket u$.
	A fourth level $\ket o$ accounts for decay to these extra levels, such that $\gamma = \gamma_g+\gamma_u+\gamma_o$ and $\beta =\frac{\gamma_g}{\gamma_g+\gamma_o}\zeta^2$. 
	\item For the calculation of the upper bound on $P^{pure}_S$, reexcitation events are excluded by setting $\gamma_u = 0$, so $\beta =\frac{\gamma_g}{\gamma}\zeta^2$.
\end{itemize}

\section{CHARACTERISATION OF EXPERIMENTAL PARAMETERS} \label{sec:exp_params}
\subsection{Cavity parameters}
\label{sec:cavpara}

In this section we provide details on the cavity and its characterisation, with key parameters summarised in Table \ref{table:cav}.

\paragraph*{Waist}
The cavity waist $w_0 = 12.31(8)~\upmu$m is calculated via $w_0=\sqrt{\lambda  \sqrt{l (2 R_C-l)}/(2 \pi) }$ from a measurement of the cavity length $l=19.906(3)$ mm and mirror radii of curvature $R_C = 9.984(7)$ mm at the wavelength $\lambda = 854$ nm.
The corresponding cavity stability parameter is $1 - l/{R_C} = 0.994(2)$ \cite{siegman86}.
The cavity length $l$ was determined from the cavity's free spectral range $FSR=c/(2l)=7530.3(2)$ MHz, with $c$ the vacuum speed of light.
The value for $R_C$ was extracted from additional measurement of the frequency spacing of the TEM$_{mn}$ modes \cite{siegman86}.
Identical $R_C$ for both mirrors are assumed.
For the measurements of $FSR$ and frequency spacings, a signal generator and fiber electro-optic-modulator (EOM) were used to modulate side bands of the respective frequency onto a laser field transmitted through the cavity.

\paragraph*{Transmission and losses}
At a wavelength of 854 nm, the finesse of the TEM$_{00}$ mode used for photon generation is $\mathcal{F} = \frac{2\pi}{\mathcal{L}} = 54(1)\times 10^3$, with the total cavity losses $\mathcal{L} = T_2 + \alpha_{loss} = 116(2) $ ppm, determined from measurements of the cavity ringdown time $\tau_\mathrm{C} = \frac {\mathcal{F}}{\pi}\cdot \frac l{c}$ using laser light. 
The cavity decay rate is given by $2\kappa = 1/\tau_C =  2\pi \times 140(3)$ kHz. 

By measuring the fraction of transmitted and reflected power of laser light at both sides of the cavity, $T_2$ and $\alpha_{loss}$ could be determined using the model of Ref. \cite{Hood2001}.
The results can be found in Table \ref{table:cav}.
Our $\alpha_{loss}=26(4)$ ppm has contributions from scattering and absorption from the mirror substrates and coatings, loss caused by imperfect alignment of the mirrors, and transmission through the other mirror with $T_1 = 2.9(4)$ ppm.
The super-polishing of the surfaces of the mirror substrates was performed by Perkins Precision, Boulder, achieving an RMS roughness of 1.0(2) \AA~ and 1.5~ppm scattering losses per mirror (as measured by them on a test piece from the batch). 
Mirror coating via ion beam sputtering was performed by Advanced Thin Films, leading to our mirror pair with combined scattering and absorption losses of 10(4)~ppm and a corresponding cavity finesse of 61(1)$\times 10^3$.
This was measured in our near-concentric configuration, but out of vacuum with each mirror mounted on a multi-axes alignment stage.

Assembling the cavity involved multiple steps to glue the mirrors into position.
In a first step the mirrors were glued (EpoTek353ND-HYB-HV) into metal tubes. 
The metal tubes were glued to piezos (Noliac shear piezos stacks) for cavity-length stabilisation, in turn glued (Masterbond EP21TCHT-1) to metal piezo-holders.
The last step consisted of gluing (Masterbond EP21TCHT-1) the piezo holders to a rigid metal spacer spanning the cavity length (involving a heat cure at 110$^\circ$C): the only tunable dimension once glued is the cavity length. 
After fully assembling the cavity out of vacuum, a finesse of 59(1)$\times 10^3$ was measured, which corresponds to an increase in $\alpha_{loss}$ of 4(2) ppm.

Integrating the cavity with the ion trap into the vacuum chamber (involving a two-week vacuum bake at 80$^\circ$C) was achieved without any measurable drop in finesse. 
After the cavity had been in vacuum for a few weeks, the finesse dropped to 54(1)$\times10^3$, corresponding to a further increase in $\alpha_{loss}$ of 9(2) ppm.
The reason for this is likely some misalignment of the cavity, the cause of which is unclear.
After that no further change of the finesse was registered.

\paragraph*{Frequency stability}
The cavity length is stabilised via the Pound-Drever-Hall (PDH) method \cite{Black2001} to a laser at 806 nm.
\textcolor{black}{
Both the 806~nm laser and the fundamental (786~nm) of the frequency-doubled drive laser are stabilised to a common in-vacuum reference cavity in order to minimise relative frequency drifts between the two arms of the CMRT.
An absolute drift rate of 720~Hz/hr of the reference cavity was observed in Ref. \cite{Hainzer2018}, which translates to a relative drift rate of the same amount between the 806~nm laser, used to lock the cavity around the ion (``ion cavity"), and drive laser after frequency doubling.
The half width of the CMRT is on the order of 100~kHz.
The drift of the reference cavity therefore causes the two arms of the CMRT (ion cavity and drive laser) to move off Raman resonance by a half linewidth in a time of about 140 hours: significantly longer than the timescale of the experiments presented here. 
}

The 806 nm laser enters the ion cavity through the mirror with transmission $T_1$.
The 806 nm wavelength lies far from any transition in $^{40}$Ca$^+$ to minimise AC Stark shifts on the atomic transitions. Furthermore, the cavity is locked to a TEM$_{01}$ mode and the ion sits in the central intensity minimum to further minimise AC Stark shifts. 
From the amplitude of the locked-cavity error signal we estimate a cavity-lock jitter of $\pm 10(2)$ kHz, 
\textcolor{black}{
	introducing phase noise and effectively broadening the CMRT. This cavity jitter is included in the form of a collapse operator onto the $\ket{D_{J{=}5/2}}$ manifold
}
in the numerical simulations presented in Secs. \ref{sec:exp_res}, \ref{sec:ent} and \ref{sec:mult_phot}.

\paragraph*{Cavity positioning}
The cavity is mounted on a system of piezo-based translation stages for 3D positioning of the cavity with respect to the trap (Attocube, ANPx311/RES/UHV for translation in the vertical dimension and along the cavity axis, ANPx321/RES/UHV for horizontal translation perpendicular to cavity axis).
The stages have integrated resistive position encoders with a specified repeatability of 1-2 $\upmu$m and can be operated in two positioning modes.
For coarse positioning, a saw-tooth signal with an amplitude of tens of Volts applied to a piezo stage effectuates a stick-slip motion with few-micron step size. 
This mode, with a range of a few mm, was used to align the cavity waist with the ion.
For fine positioning, applying a DC voltage allows for sub-nm displacement. 
By fine tuning of the cavity position along the cavity axis, the ion is placed in the center of an anti-node of the cavity standing wave before each experiment. A voltage of up to $\pm$9 V from a battery is applied to the corresponding positioner while maximising the photon count rate.

The relative position of the waist to the ion was determined by probing the intra-cavity field of a laser with an ion.
The ion was translated perpendicular to the cavity axis through a cavity mode and the position-dependent AC-Stark shift on the $\ket{4^{2} S_{J{=}1/2},m_j{=}-1/2} \leftrightarrow \ket{4^{2} D_{J{=}5/2},m_j{=}-5/2}$ transition was measured with a narrow-linewidth (on the order of 1 Hz) laser at 729 nm.
The position of the ion with respect to the cavity waist could then be extracted from a fit of the resulting fringe pattern of the standing-wave with a model taking into account the Hermite-Gauss mode pattern of the cavity mode.

\begin{table}[t]
	\centering
	\begin{tabular}{|c|c|c|c|} 
		\hline
		$w_0$ ($\upmu$m) & $\alpha_{loss}$ (ppm) & $T_2$ (ppm) & $\kappa/2\pi$ (kHz) \\ [0.5ex] 
		\hline\hline
		12.31(8) & 26(4) & 90(4) & 70(2) \\ [0.5ex] 
		\hline
	\end{tabular}
	\caption{Key cavity parameters. $w_0$: cavity waist. $\alpha_{loss}$: unwanted cavity loss. $T_2$: output mirror transmission. $\kappa$: cavity decay rate (half width at half maximum).}
	\label{table:cav}
\end{table}

\subsection{Photon path and detectors}\label{sec:path}

After exiting the cavity, photons pass the following optical elements: an in-vacuum collimating lens, a vacuum viewport, a mirror, a lens and three optical filters.
Combined, these elements have a transmission of $P_{el} = 0.97(1)$. 
Photons are then coupled to a single-mode fiber for which a maximum coupling efficiency of $0.82(3)$ was determined with classical light coupled through the cavity (thus matching the single-photon's path), measuring the power at the fiber input and output with photo diodes.
The error bars are due to fluctuations of the laser light intensity transmitted through the cavity.
When opening and closing the $\mu$-metal shield surrounding the vacuum chamber, optical elements and fiber in-coupling stage, slight additional drifts lead us to estimate the fiber-coupling efficiency to be $P_{fc} = 0.81(3)$.
Photons are detected with a fiber-coupled superconducting nano-wire single-photon detector (SNSPD) from Scontel which has an efficiency $P_{det} = 0.87(2)$ and free-running dark counts of 0.3(1) per second at 854 nm.

Only for the ion-photon entanglement experiment (Sec. \ref{sec:ent}) a half- and quarter-wave plate as well as polarising beam splitter (PBS) are added into the photon path, introducing another 1\% of loss [i.e. $P_{el}$ is reduced by 0.01 over its previous value to 0.96(1)].
The reflected port of the PBS is coupled to a second fiber with $P_{fc} = 0.80(3)$ and the corresponding second detector has $P_{det} = 0.88(2)$ with free-running dark counts of 0.5(1) per second.

The detector efficiencies were calibrated during installation of the system by the manufacturer, using classical light, powermeter and calibrated attenuators. 
The calibration was cross-checked by comparison with (independently calibrated) avalanche photo-diodes (APDs) from Laser-Components. 


\subsection{Drive-laser Rabi frequency and ion temperature} \label{sec:calibration}

In the case of a single-frequency drive-laser, its Rabi frequency $\Omega$ is determined by measuring the AC-Stark shift $\delta_{AC}$ that the laser induces in the frequency of the CMRT: $\Omega = \sqrt{4\delta_{AC}\Delta}$.
The detuning $\Delta/2\pi = -403(5)$ MHz, measured with a wavemeter, is the detuning of the drive laser from the $\ket u \leftrightarrow \ket e$ transition. 

The frequency difference in the bichromatic drive field in Sec. \ref{sec:ent} (ion-photon entanglement) is equal to the Zeeman splitting between the states $\ket{g_1}$ and $\ket{g_2}$, found by performing spectroscopy on the 729 nm transition.
The total Rabi frequency $\Omega = \sqrt{\Omega_1^2 + \Omega_2^2}$ is approximated with the same formula as for the single-frequency case, taking the same detuning $\Delta$ for both components.

For each experiment presented in the paper the ion was cooled close to the ground state of motion.
Via resolved side-band cooling on the 729 nm transition we achieve a mean phonon number \textcolor{black}{$\bar n = 0.5(2)$} in each motional mode.
The  mean phonon number $\bar n$ is determined via Rabi flops on the 729 nm ($S_{1/2} \leftrightarrow D_{5/2}$) transition and fitting the observed dependence of the excitation probability on the 729 nm pulse length with a model that takes into account the ion temperature \cite{Roos}.  
Rabi flops are performed with two different 729 nm laser beam directions, allowing the temperature in different motional modes to be distinguished.

\section{IMPROVEMENTS OVER THE EXPERIMENT OF REF. \cite{Krutyanskiy2019}} \label{sec:upgrades}
In Ref. \cite{Krutyanskiy2019}, our ion-cavity system was used to achieve $P_{tot} = 0.08(1)$ for ion-entangled photons.  
We now summarise four key technical advances, made over the system in Ref. \cite{Krutyanskiy2019}, that enabled the performance presented in the present manuscript.
First, using the method described in Sec. \ref{sec:cavpara}, the ion was discovered to be approximately 0.5 mm away from the position of the cavity waist along the cavity axis. By correcting this imperfection, we estimate a reduction in the cavity effective mode area $A_{e\!f\!f}$ at the point of the ion by a factor of two and corresponding increase in cooperativity $C$. 
Second, 3D ground state cooling of the ion’s motional state was carried out, compared with only Doppler cooling in Ref. \cite{Krutyanskiy2019}. 
After Doppler cooling in our system, mean phonon numbers of 11(2) and 8(2) are determined in the axial and radial modes, respectively. 
The dominant effect of being outside the ground state on cavity-photon generation in our system is due to changes in the coupling of the drive laser to the ion. 
Specifically, for a motional mode with phonon number $n$, the drive laser Rabi frequency in the CMRT can be approximated by $\Omega_n \approx \Omega(1-\eta^2 n)$, where $\eta$ is the Lamb-Dicke parameter. 
The approximation holds for $\eta^2(2n+1)\ll 1$ (the Lamb-Dicke regime). For the full expression without approximation, see, e.g., Ref. \cite{Wineland1998}. 
Outside of the ground state, the coupling of our 393~nm drive-laser to the axial mode ($\eta = 0.13$) causes the most significant reduction of the Rabi frequency $\Omega_n$. 
At the same time, the spontaneous decay rate remains effectively constant at Doppler-cooled temperatures and below, leading to a reduced maximum efficiency for cavity photon production. 
Third, the Rabi frequency of the drive laser used to achieve the highest efficiency in the present work ($\Omega/2\pi=14$ MHz) is significantly lower than the value used in Ref. \cite{Krutyanskiy2019} ($\Omega/2\pi \approx 27$ MHz), thereby further reducing spontaneous scattering to the $\ket{g,0}, \ket{o,0}$ states. 
Fourth, the detection path efficiency $P_{path}$ was increased: the single-mode fiber coupling was improved from 0.5(1) to 0.82(5), and the photon detector efficiency was improved from 0.4 to 0.87(2) by moving from avalanche photo-diodes to superconducting nanowire detectors.

\section{CHOICE OF $T_2$} \label{sec:T2}

\begin{figure}[t]
	\vspace{-6mm}
	\begin{center}
		\includegraphics[width=\columnwidth]{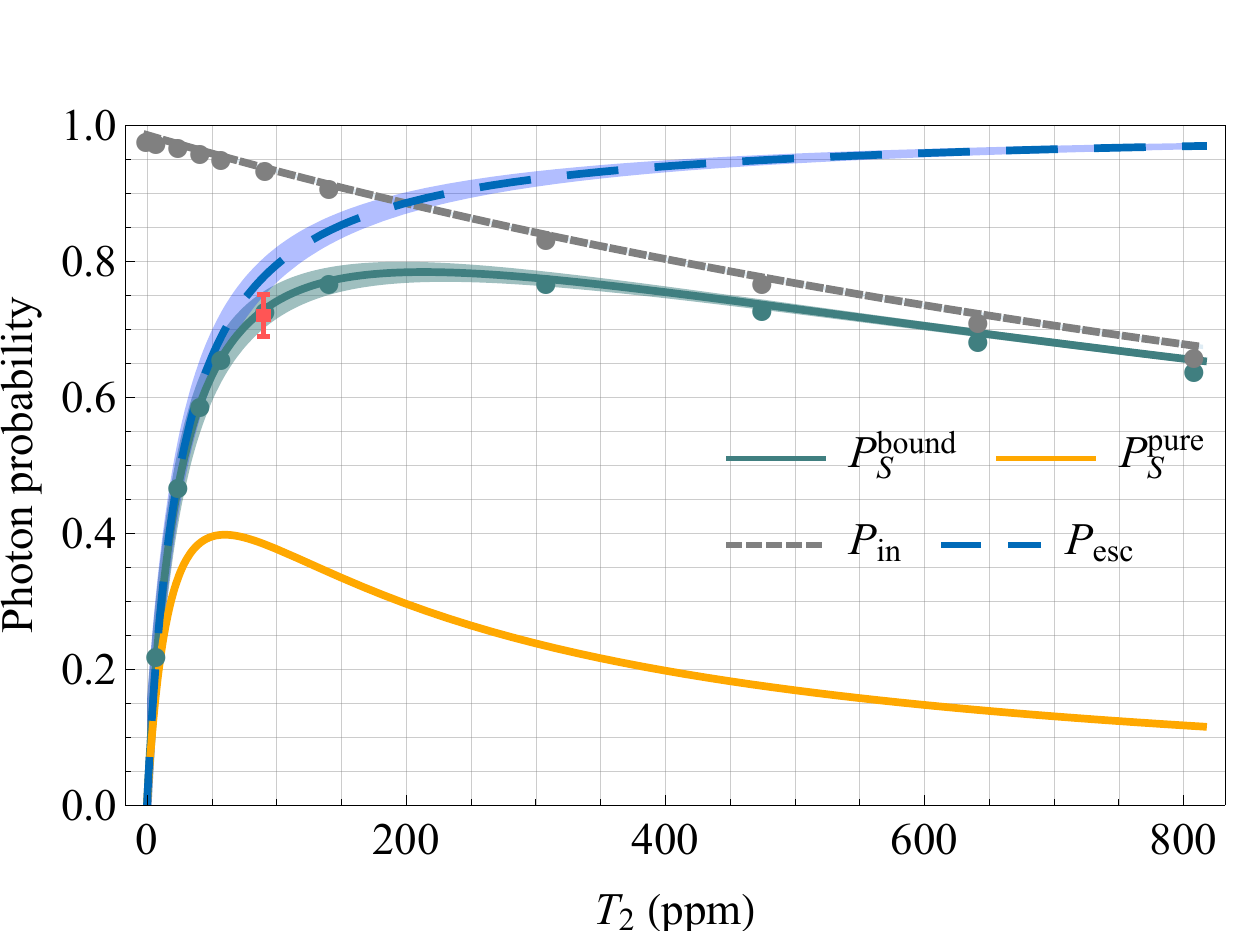}
		\vspace{0mm}
		\caption{
			\textbf{Predicted photon probabilities in our system as a function of the output mirror transmission $T_2$.} 
			The probability for obtaining a photon in the output mode of the cavity $P_S = P_{esc} P_{in}$ results from the product of the probability for photon emission into the cavity $P_{in}$ and escape probability $P_{esc}$.
			Lines: analytical calculations, where $P_S^{bound}$ and $P^{pure}_S$ are calculated via Eq. \eqref{eq:PS} ($j = 0$ for $P^{pure}_S$) for the experimental configuration shown in Fig. \ref{fig:exp} and the parameters given in Sec. \ref{sec:PSexp}.
			Shaded areas around $P_S^{bound}$ and $P_{esc}$ represent the uncertainties in the measured cavity parameters.
			Red dot with error bars: experimentally measured efficiency $P^{exp}_S$; 
			error bars are due to the uncertainty in the efficiency of detector and photon path.
			Dots: numerical simulations, taking into account our particular drive scheme introduced in Sec. \ref{sec:exp_det}. 
			The numerically calculated efficiencies converge towards the analytic calculations for $\Omega/\Delta \ll 1$, where $\Omega$ is the Rabi frequency of the drive laser and $\Delta$ the common detuning of cavity and drive laser from the excited state. 
			Slight deviations for higher $T_2$ are due to the finite evolution time used in the numerical simulations.
		}
		\label{fig:PvsT2}
		\vspace{-6mm}
	\end{center}
\end{figure}

The optimised upper bound $P_S^{opt}$, given by Eq. \eqref{eq:PSopt}, requires the optimum output-mirror transmission $T^{opt}_2$ calculated via Eq. \eqref{eq:T2opt}. 
The following calculations hold for our experimental configuration shown in Fig. \ref{fig:exp}, with relevant parameters summarised in Sec. \ref{sec:PSexp}.
Allowing for a different value of our outcoupling mirror transmission, without changing $\alpha_{loss}$, $w_0$ or the propagation direction of the drive laser, would only lead to a relatively small increase in $P_S^{bound}$: a value of $T_2^{opt}=216$ ppm would increase $P_S^{bound}$ from 0.73 to 0.78 at the expense of reducing $P_S^{pure}$. 
In general there is a trade-off here.
The probability $P^{pure}_S$ for obtaining a transform-limited photon (i.e without prior decay to the initial state, see section \ref{sec:model}), has an upper bound that requires the optimum output-mirror transmission $T^{opt.pure}_2 = 61$ ppm for our experiment, resulting in $P^{pure}_S \le 0.39$.
With our $T_2 = 90(4)$ ppm we get close the the upper bounds of both $P_S$ and $P^{pure}_S$, as can be seen in Fig. \ref{fig:PvsT2}.

\section{POSSIBLE EFFICIENCY IMPROVEMENTS IN OUR EXISTING SETUP} \label{sec:improvements2}
\begin{figure}[t]
	\vspace{0mm}
	\begin{center}
		\includegraphics[width=\columnwidth]{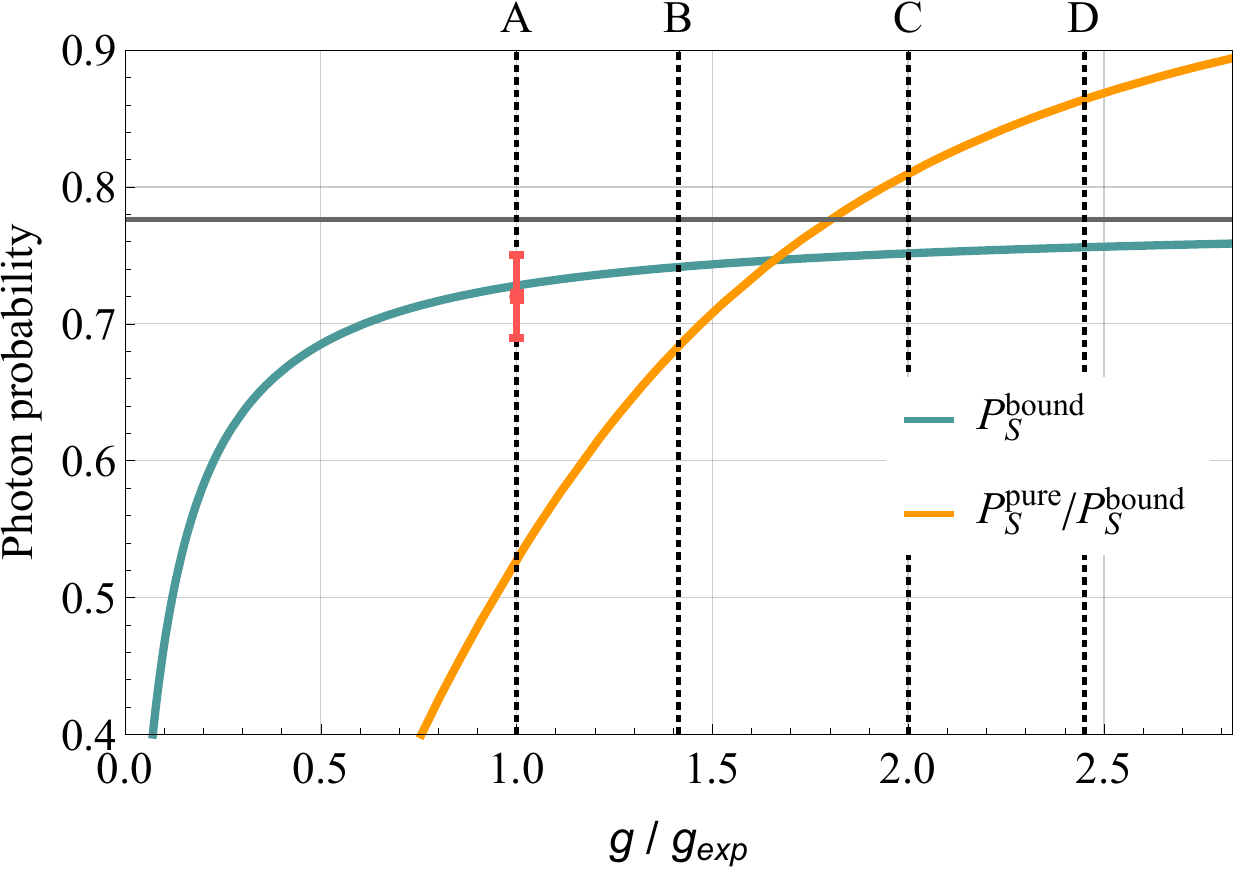}
		\vspace{0mm}
		\caption{
			\textbf{Possible efficiency improvements in our existing ion-cavity system.} 
			Theoretical predictions for $P^{bound}_S$ and $P^{pure}_S/P^{bound}_S$, calculated via Eq. \eqref{eq:PS}, with $j=0$ for $P^{pure}_S$, when increasing the ion-cavity coupling strength $g$ over its current value $g_{exp}$.
			Labels A-D represent possible schemes in our system.
			Theoretical predictions use mean values for experimental parameters.
			Scheme A corresponds to the current experimental configuration (Fig. \ref{fig:exp}), where $\zeta = \sqrt{0.5}$.
			Scheme B corresponds to changing the orientation of the drive beam and magnetic field axis to be parallel to the cavity axis, such that $\zeta = 1$.
			Schemes C and D correspond to coupling two and three ions, respectively, in superradiant entangled states to the cavity, in addition to scheme B.
			Red dot with error bars: measured efficiency for a drive field with $\Omega = 14$~MHz (Sec. \ref{sec:exp_res}).
			Horizontal grey line: photon escape probability through the cavity output mirror $P_{esc}=T_2/(T_2 + \alpha_{loss})=0.78$.
		}
		\label{fig:Pvsg}
		\vspace{-6mm}
	\end{center}
\end{figure}

In this section, schemes are presented
to enhance performance in our system without changing properties of the optical cavity. 
These schemes seek to increase the ion-cavity coupling strength $g$ which in turn would increase the cooperativity (Eq. \ref{eq:coop2}).
Scheme A is our experimental benchmark: the  configuration in Fig. \ref{fig:exp}, where $g_{exp}/2\pi = 0.88(1)$ MHz. 
In scheme B, the drive laser propagates parallel to the magnetic field axis and cavity axis, achieving a larger polarisation projection $\zeta = 1$ onto the dipole of the $\ket g \leftrightarrow \ket e$ transition, compared to Scheme A.
Schemes C and D correspond to coupling two and three ions, respectively, in superradiant entangled states to the cavity, in addition to scheme B. 
Enhanced cavity coupling via superradiance has been demonstrated for two entangled ions \cite{Casabone2015} and it is feasible to near-maximally couple three ions to our cavity. 
Fig. \ref{fig:Pvsg} compares the performance of schemes A-D, as a function of the normalised ion-cavity coupling strength $g/g_{exp}$. 
No significant improvement in $P^{bound}_S$ is evident: the value achieved in scheme A is already close to saturation ($P_{esc}$). In contrast, schemes B - D would enable significant improvements in $P_S^{pure}/P^{bound}_S$: the fraction of photons in the cavity output mode generated without prior spontaneous decay of the emitter can be significantly increased. 

\section{EXPERIMENTAL SEQUENCE} \label{sec:sequence}
\begin{figure*}[h!]
	\begin{center}
		\includegraphics[width=1\textwidth]{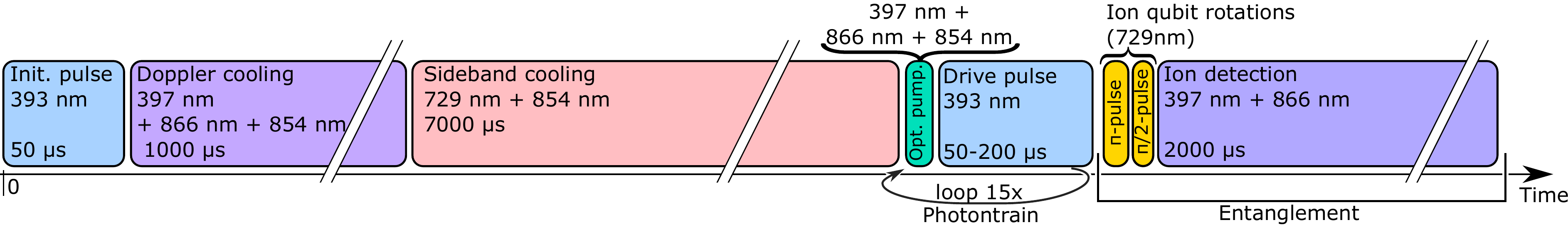}
		\vspace{0mm}
		\caption{
			\textbf{Laser pulse sequences for the experiments presented in the paper}.
			An initialisation laser-pulse for intensity stabilisation of the drive laser is followed by Doppler cooling. 
			Subsequently, the ion is cooled close to the ground state of each of its three motional modes and prepared in the initial state $\ket u =\ket{S_{J=1/2},m_j=-1/2}$ via optical pumping ($10~\upmu$s of $\sigma^-$-polarised 397 nm light combined with 866 nm; 854 nm for repumping from the $D_{J=5/2}$-manifold). 
			For the multi-photon experiment (Sec. \ref{sec:mult_phot}) the drive laser pulse, preceded by optical pumping, is repeated 15 times ('photontrain'). 
			In the case of ion-photon entanglement (Sec. \ref{sec:ent}) the drive laser pulse is bichromatic and ion-qubit rotations as well as ion state detection are performed. 
			See Fig. \ref{fig:full_level} for corresponding atomic transitions and Sec. \ref{sec:state_characterisation} for details on the lasers pulses for reconstructing the ion-photon entanglement. 
		}
		\label{fig:seq}
		\vspace{-6mm}
	\end{center}
\end{figure*}

The laser-pulse sequences for generating single photons (Sec. \ref{sec:exp_res}), ion-photon entanglement (Sec. \ref{sec:ent}) and multiple sequential photons (Sec. \ref{sec:mult_phot}) is shown in Fig. \ref{fig:seq}.

\section{RECONSTRUCTION OF THE ION-PHOTON ENTANGLED STATE} \label{sec:state_characterisation}

To reconstruct the ion-photon state, a full state tomography of the two-qubit system is performed.
On the photon polarisation-qubit side, the state is projected to one of 6 states (horizontal, vertical, diagonal, anti-diagonal, right circular and left circular) by waveplates and polariser. 
This is equivalent to performing projective measurements in three bases described by the Pauli spin-1/2 operators. For example, horizontal and vertical are the eigenstates of the Pauli $\sigma_z$ operator. 

To perform an ion qubit measurement the $\ket{g_1} = \ket{D_{J=5/2},m_j=-5/2}$ electron population is first mapped to the $\ket u = \ket{S_{J=1/2},m_j=-1/2}$ state via a 729 nm $\pi$-pulse. 
That is, the D-manifold qubit is mapped into an optical qubit (with logical states $\ket u$ and $\ket{g_2} = \ket{D_{J=5/2},m_j=-3/2}$). 
In order to measure which of these states the electron is in, the standard electron shelving technique is used: in the case of the $\ket u$-state outcome, 397 nm photons from the ion are collected with free-space optics for a detection time of 2000 $\upmu$s. 
This is sufficient to distinguish bright (scattering) and dark (non-scattering) ions with an error of less than 1\%. 
The aforementioned process implements a projective measurement into the eigenstates of the $\sigma_z$ basis. 
For measurements in other bases e.g $\sigma_x$ ($\sigma_y$), as required for full quantum state tomography, an additional 729 nm $\pi/2$-pulse on the $\ket u \leftrightarrow \ket{g_2}$ transition with a 0 ($\pi/2$) phase is applied after the $\pi$ pulse and before the 397~nm pulse, to rotate the ion-qubit measurement basis. 

For each of the 9 possible joint measurement bases (choice of photon basis and ion basis), the numbers of events corresponding to one of the four possible outcomes of these 2-qubit measurements are recorded (there is therefore a total of 36 possible outcomes).
We then divide the number of events recorded for each outcome by the total number of events recorded for the given basis (divide each number by the sum of four) and thus obtain estimates of the outcome probabilities.   
These probabilities are used to reconstruct the 2-qubit density matrix by linear search with subsequent maximum likelihood method \cite{Jezek2003}.

For statistical analysis (determining error bars in quantities derived from the reconstructed density matrix), the Monte-Carlo approach was used. 
Briefly, we numerically generate M = 200 sets of 36 event numbers with Poissonian distribution and mean value equal to the experimental value for each of the 36 possible outcomes. 
From these simulated event numbers we derive simulated outcome probabilities, the same way as we do for the experimental counts.
Then we reconstruct M density matrices for this simulated data and for each one we calculate the quantities of interest (fidelity, purity). The error bars given in the main text represent one standard deviation in the widths of the distributions of these quantities over M simulated data sets.  

The limit of the fidelity due to background counts was obtained by adding unpolarised background detection events to the expected counts for a $\Phi^+$ Bell state.
Background counts were determined by counting the photon detection events in a 60 $\upmu$s window outside the Raman pulse for 45000 repetitions.



\begin{thebibliography}{69}%
	\makeatletter
	\providecommand \@ifxundefined [1]{%
		\@ifx{#1\undefined}
	}%
	\providecommand \@ifnum [1]{%
		\ifnum #1\expandafter \@firstoftwo
		\else \expandafter \@secondoftwo
		\fi
	}%
	\providecommand \@ifx [1]{%
		\ifx #1\expandafter \@firstoftwo
		\else \expandafter \@secondoftwo
		\fi
	}%
	\providecommand \natexlab [1]{#1}%
	\providecommand \enquote  [1]{``#1''}%
	\providecommand \bibnamefont  [1]{#1}%
	\providecommand \bibfnamefont [1]{#1}%
	\providecommand \citenamefont [1]{#1}%
	\providecommand \href@noop [0]{\@secondoftwo}%
	\providecommand \href [0]{\begingroup \@sanitize@url \@href}%
	\providecommand \@href[1]{\@@startlink{#1}\@@href}%
	\providecommand \@@href[1]{\endgroup#1\@@endlink}%
	\providecommand \@sanitize@url [0]{\catcode `\\12\catcode `\$12\catcode
		`\&12\catcode `\#12\catcode `\^12\catcode `\_12\catcode `\%12\relax}%
	\providecommand \@@startlink[1]{}%
	\providecommand \@@endlink[0]{}%
	\providecommand \url  [0]{\begingroup\@sanitize@url \@url }%
	\providecommand \@url [1]{\endgroup\@href {#1}{\urlprefix }}%
	\providecommand \urlprefix  [0]{URL }%
	\providecommand \Eprint [0]{\href }%
	\providecommand \doibase [0]{https://doi.org/}%
	\providecommand \selectlanguage [0]{\@gobble}%
	\providecommand \bibinfo  [0]{\@secondoftwo}%
	\providecommand \bibfield  [0]{\@secondoftwo}%
	\providecommand \translation [1]{[#1]}%
	\providecommand \BibitemOpen [0]{}%
	\providecommand \bibitemStop [0]{}%
	\providecommand \bibitemNoStop [0]{.\EOS\space}%
	\providecommand \EOS [0]{\spacefactor3000\relax}%
	\providecommand \BibitemShut  [1]{\csname bibitem#1\endcsname}%
	\let\auto@bib@innerbib\@empty
	\bibitem [{\citenamefont {Solomon}\ \emph {et~al.}(2013)\citenamefont
		{Solomon}, \citenamefont {Santori},\ and\ \citenamefont
		{Kuhn}}]{Solomon2013}%
	\BibitemOpen
	\bibfield  {author} {\bibinfo {author} {\bibfnamefont {G.~S.}\ \bibnamefont
			{Solomon}}, \bibinfo {author} {\bibfnamefont {C.}~\bibnamefont {Santori}},\
		and\ \bibinfo {author} {\bibfnamefont {A.}~\bibnamefont {Kuhn}},\ }\bibfield
	{title} {\bibinfo {title} {{Single Emitters in Isolated Quantum Systems}},\
	}in\ \href {https://doi.org/10.1016/B978-0-12-387695-9.00013-5} {\emph
		{\bibinfo {booktitle} {Experimental Methods in the Physical Sciences}}},\
	Vol.~\bibinfo {volume} {45}\ (\bibinfo  {publisher} {Elsevier Inc.},\
	\bibinfo {year} {2013})\ pp.\ \bibinfo {pages} {467--539}\BibitemShut
	{NoStop}%
	\bibitem [{\citenamefont {Heshami}\ \emph {et~al.}(2016)\citenamefont
		{Heshami}, \citenamefont {England}, \citenamefont {Humphreys}, \citenamefont
		{Bustard}, \citenamefont {Acosta}, \citenamefont {Nunn},\ and\ \citenamefont
		{Sussman}}]{Heshami2016}%
	\BibitemOpen
	\bibfield  {author} {\bibinfo {author} {\bibfnamefont {K.}~\bibnamefont
			{Heshami}}, \bibinfo {author} {\bibfnamefont {D.~G.}\ \bibnamefont
			{England}}, \bibinfo {author} {\bibfnamefont {P.~C.}\ \bibnamefont
			{Humphreys}}, \bibinfo {author} {\bibfnamefont {P.~J.}\ \bibnamefont
			{Bustard}}, \bibinfo {author} {\bibfnamefont {V.~M.}\ \bibnamefont {Acosta}},
		\bibinfo {author} {\bibfnamefont {J.}~\bibnamefont {Nunn}},\ and\ \bibinfo
		{author} {\bibfnamefont {B.~J.}\ \bibnamefont {Sussman}},\ }\bibfield
	{title} {\bibinfo {title} {{Quantum memories: emerging applications and
				recent advances}},\ }\href {https://doi.org/10.1080/09500340.2016.1148212}
	{\bibfield  {journal} {\bibinfo  {journal} {J. Mod. Opt.}\ }\textbf {\bibinfo
			{volume} {63}},\ \bibinfo {pages} {2005} (\bibinfo {year}
		{2016})}\BibitemShut {NoStop}%
	\bibitem [{\citenamefont {Monroe}\ \emph {et~al.}(2014)\citenamefont {Monroe},
		\citenamefont {Raussendorf}, \citenamefont {Ruthven}, \citenamefont {Brown},
		\citenamefont {Maunz}, \citenamefont {Duan},\ and\ \citenamefont
		{Kim}}]{Monroe2014}%
	\BibitemOpen
	\bibfield  {author} {\bibinfo {author} {\bibfnamefont {C.}~\bibnamefont
			{Monroe}}, \bibinfo {author} {\bibfnamefont {R.}~\bibnamefont {Raussendorf}},
		\bibinfo {author} {\bibfnamefont {A.}~\bibnamefont {Ruthven}}, \bibinfo
		{author} {\bibfnamefont {K.~R.}\ \bibnamefont {Brown}}, \bibinfo {author}
		{\bibfnamefont {P.}~\bibnamefont {Maunz}}, \bibinfo {author} {\bibfnamefont
			{L.-M.}\ \bibnamefont {Duan}},\ and\ \bibinfo {author} {\bibfnamefont
			{J.}~\bibnamefont {Kim}},\ }\bibfield  {title} {\bibinfo {title}
		{{Large-scale modular quantum-computer architecture with atomic memory and
				photonic interconnects}},\ }\href
	{https://doi.org/10.1103/PhysRevA.89.022317} {\bibfield  {journal} {\bibinfo
			{journal} {Phys. Rev. A}\ }\textbf {\bibinfo {volume} {89}},\ \bibinfo
		{pages} {022317} (\bibinfo {year} {2014})}\BibitemShut {NoStop}%
	\bibitem [{\citenamefont {Kimble}(2008)}]{Kimble2008}%
	\BibitemOpen
	\bibfield  {author} {\bibinfo {author} {\bibfnamefont {H.~J.}\ \bibnamefont
			{Kimble}},\ }\bibfield  {title} {\bibinfo {title} {{The quantum internet}},\
	}\href {https://doi.org/10.1038/nature07127} {\bibfield  {journal} {\bibinfo
			{journal} {Nature}\ }\textbf {\bibinfo {volume} {453}},\ \bibinfo {pages}
		{1023} (\bibinfo {year} {2008})}\BibitemShut {NoStop}%
	\bibitem [{\citenamefont {Wehner}\ \emph {et~al.}(2018)\citenamefont {Wehner},
		\citenamefont {Elkouss},\ and\ \citenamefont {Hanson}}]{Wehner2018}%
	\BibitemOpen
	\bibfield  {author} {\bibinfo {author} {\bibfnamefont {S.}~\bibnamefont
			{Wehner}}, \bibinfo {author} {\bibfnamefont {D.}~\bibnamefont {Elkouss}},\
		and\ \bibinfo {author} {\bibfnamefont {R.}~\bibnamefont {Hanson}},\
	}\bibfield  {title} {\bibinfo {title} {{Quantum internet: A vision for the
				road ahead}},\ }\href {https://doi.org/10.1126/science.aam9288} {\bibfield
		{journal} {\bibinfo  {journal} {Science}\ }\textbf {\bibinfo {volume}
			{362}},\ \bibinfo {pages} {eaam9288} (\bibinfo {year} {2018})}\BibitemShut
	{NoStop}%
	\bibitem [{\citenamefont {Gisin}\ \emph {et~al.}(2002)\citenamefont {Gisin},
		\citenamefont {Ribordy}, \citenamefont {Tittel},\ and\ \citenamefont
		{Zbinden}}]{Gisin2002}%
	\BibitemOpen
	\bibfield  {author} {\bibinfo {author} {\bibfnamefont {N.}~\bibnamefont
			{Gisin}}, \bibinfo {author} {\bibfnamefont {G.}~\bibnamefont {Ribordy}},
		\bibinfo {author} {\bibfnamefont {W.}~\bibnamefont {Tittel}},\ and\ \bibinfo
		{author} {\bibfnamefont {H.}~\bibnamefont {Zbinden}},\ }\bibfield  {title}
	{\bibinfo {title} {{Quantum cryptography}},\ }\href
	{https://doi.org/10.1103/RevModPhys.74.145} {\bibfield  {journal} {\bibinfo
			{journal} {Rev. Mod. Phys.}\ }\textbf {\bibinfo {volume} {74}},\ \bibinfo
		{pages} {145} (\bibinfo {year} {2002})}\BibitemShut {NoStop}%
	\bibitem [{\citenamefont {Sekatski}\ \emph {et~al.}(2020)\citenamefont
		{Sekatski}, \citenamefont {W{\"{o}}lk},\ and\ \citenamefont
		{D{\"{u}}r}}]{Sekatski2020}%
	\BibitemOpen
	\bibfield  {author} {\bibinfo {author} {\bibfnamefont {P.}~\bibnamefont
			{Sekatski}}, \bibinfo {author} {\bibfnamefont {S.}~\bibnamefont
			{W{\"{o}}lk}},\ and\ \bibinfo {author} {\bibfnamefont {W.}~\bibnamefont
			{D{\"{u}}r}},\ }\bibfield  {title} {\bibinfo {title} {{Optimal distributed
				sensing in noisy environments}},\ }\href
	{https://doi.org/10.1103/PhysRevResearch.2.023052} {\bibfield  {journal}
		{\bibinfo  {journal} {Phys. Rev. Research}\ }\textbf {\bibinfo {volume}
			{2}},\ \bibinfo {pages} {023052} (\bibinfo {year} {2020})}\BibitemShut
	{NoStop}%
	\bibitem [{\citenamefont {K{\'{o}}m{\'{a}}r}\ \emph {et~al.}(2014)\citenamefont
		{K{\'{o}}m{\'{a}}r}, \citenamefont {Kessler}, \citenamefont {Bishof},
		\citenamefont {Jiang}, \citenamefont {S{\o}rensen}, \citenamefont {Ye},\ and\
		\citenamefont {Lukin}}]{Komar2014}%
	\BibitemOpen
	\bibfield  {author} {\bibinfo {author} {\bibfnamefont {P.}~\bibnamefont
			{K{\'{o}}m{\'{a}}r}}, \bibinfo {author} {\bibfnamefont {E.~M.}\ \bibnamefont
			{Kessler}}, \bibinfo {author} {\bibfnamefont {M.}~\bibnamefont {Bishof}},
		\bibinfo {author} {\bibfnamefont {L.}~\bibnamefont {Jiang}}, \bibinfo
		{author} {\bibfnamefont {A.~S.}\ \bibnamefont {S{\o}rensen}}, \bibinfo
		{author} {\bibfnamefont {J.}~\bibnamefont {Ye}},\ and\ \bibinfo {author}
		{\bibfnamefont {M.~D.}\ \bibnamefont {Lukin}},\ }\bibfield  {title} {\bibinfo
		{title} {{A quantum network of clocks}},\ }\href
	{https://doi.org/10.1038/nphys3000} {\bibfield  {journal} {\bibinfo
			{journal} {Nat. Phys.}\ }\textbf {\bibinfo {volume} {10}},\ \bibinfo {pages}
		{582} (\bibinfo {year} {2014})}\BibitemShut {NoStop}%
	\bibitem [{\citenamefont {Duan}\ \emph {et~al.}(2001)\citenamefont {Duan},
		\citenamefont {Lukin}, \citenamefont {Cirac},\ and\ \citenamefont
		{Zoller}}]{Duan2001}%
	\BibitemOpen
	\bibfield  {author} {\bibinfo {author} {\bibfnamefont {L.-M.}\ \bibnamefont
			{Duan}}, \bibinfo {author} {\bibfnamefont {M.~D.}\ \bibnamefont {Lukin}},
		\bibinfo {author} {\bibfnamefont {J.~I.}\ \bibnamefont {Cirac}},\ and\
		\bibinfo {author} {\bibfnamefont {P.}~\bibnamefont {Zoller}},\ }\bibfield
	{title} {\bibinfo {title} {{Long-distance quantum communication with atomic
				ensembles and linear optics}},\ }\href {https://doi.org/10.1038/35106500}
	{\bibfield  {journal} {\bibinfo  {journal} {Nature}\ }\textbf {\bibinfo
			{volume} {414}},\ \bibinfo {pages} {413} (\bibinfo {year}
		{2001})}\BibitemShut {NoStop}%
	\bibitem [{\citenamefont {Reiserer}\ and\ \citenamefont
		{Rempe}(2015)}]{Reiserer2015}%
	\BibitemOpen
	\bibfield  {author} {\bibinfo {author} {\bibfnamefont {A.}~\bibnamefont
			{Reiserer}}\ and\ \bibinfo {author} {\bibfnamefont {G.}~\bibnamefont
			{Rempe}},\ }\bibfield  {title} {\bibinfo {title} {{Cavity-based quantum
				networks with single atoms and optical photons}},\ }\href
	{https://doi.org/10.1103/RevModPhys.87.1379} {\bibfield  {journal} {\bibinfo
			{journal} {Rev. Mod. Phys.}\ }\textbf {\bibinfo {volume} {87}},\ \bibinfo
		{pages} {1379} (\bibinfo {year} {2015})}\BibitemShut {NoStop}%
	\bibitem [{\citenamefont {Junge}\ \emph {et~al.}(2013)\citenamefont {Junge},
		\citenamefont {O'Shea}, \citenamefont {Volz},\ and\ \citenamefont
		{Rauschenbeutel}}]{Junge2013}%
	\BibitemOpen
	\bibfield  {author} {\bibinfo {author} {\bibfnamefont {C.}~\bibnamefont
			{Junge}}, \bibinfo {author} {\bibfnamefont {D.}~\bibnamefont {O'Shea}},
		\bibinfo {author} {\bibfnamefont {J.}~\bibnamefont {Volz}},\ and\ \bibinfo
		{author} {\bibfnamefont {A.}~\bibnamefont {Rauschenbeutel}},\ }\bibfield
	{title} {\bibinfo {title} {{Strong Coupling between Single Atoms and
				Nontransversal Photons}},\ }\href
	{https://doi.org/10.1103/PhysRevLett.110.213604} {\bibfield  {journal}
		{\bibinfo  {journal} {Phys. Rev. Lett.}\ }\textbf {\bibinfo {volume} {110}},\
		\bibinfo {pages} {213604} (\bibinfo {year} {2013})}\BibitemShut {NoStop}%
	\bibitem [{\citenamefont {Wang}\ \emph {et~al.}(2019)\citenamefont {Wang},
		\citenamefont {He}, \citenamefont {Chung}, \citenamefont {Hu}, \citenamefont
		{Yu}, \citenamefont {Chen}, \citenamefont {Ding}, \citenamefont {Chen},
		\citenamefont {Qin}, \citenamefont {Yang}, \citenamefont {Liu}, \citenamefont
		{Duan}, \citenamefont {Li}, \citenamefont {Gerhardt}, \citenamefont
		{Winkler}, \citenamefont {Jurkat}, \citenamefont {Wang}, \citenamefont
		{Gregersen}, \citenamefont {Huo}, \citenamefont {Dai}, \citenamefont {Yu},
		\citenamefont {H{\"{o}}fling}, \citenamefont {Lu},\ and\ \citenamefont
		{Pan}}]{Wang2019a}%
	\BibitemOpen
	\bibfield  {author} {\bibinfo {author} {\bibfnamefont {H.}~\bibnamefont
			{Wang}}, \bibinfo {author} {\bibfnamefont {Y.-M.}\ \bibnamefont {He}},
		\bibinfo {author} {\bibfnamefont {T.-H.}\ \bibnamefont {Chung}}, \bibinfo
		{author} {\bibfnamefont {H.}~\bibnamefont {Hu}}, \bibinfo {author}
		{\bibfnamefont {Y.}~\bibnamefont {Yu}}, \bibinfo {author} {\bibfnamefont
			{S.}~\bibnamefont {Chen}}, \bibinfo {author} {\bibfnamefont {X.}~\bibnamefont
			{Ding}}, \bibinfo {author} {\bibfnamefont {M.-C.}\ \bibnamefont {Chen}},
		\bibinfo {author} {\bibfnamefont {J.}~\bibnamefont {Qin}}, \bibinfo {author}
		{\bibfnamefont {X.}~\bibnamefont {Yang}}, \bibinfo {author} {\bibfnamefont
			{R.-Z.}\ \bibnamefont {Liu}}, \bibinfo {author} {\bibfnamefont {Z.-C.}\
			\bibnamefont {Duan}}, \bibinfo {author} {\bibfnamefont {J.-P.}\ \bibnamefont
			{Li}}, \bibinfo {author} {\bibfnamefont {S.}~\bibnamefont {Gerhardt}},
		\bibinfo {author} {\bibfnamefont {K.}~\bibnamefont {Winkler}}, \bibinfo
		{author} {\bibfnamefont {J.}~\bibnamefont {Jurkat}}, \bibinfo {author}
		{\bibfnamefont {L.-J.}\ \bibnamefont {Wang}}, \bibinfo {author}
		{\bibfnamefont {N.}~\bibnamefont {Gregersen}}, \bibinfo {author}
		{\bibfnamefont {Y.-H.}\ \bibnamefont {Huo}}, \bibinfo {author} {\bibfnamefont
			{Q.}~\bibnamefont {Dai}}, \bibinfo {author} {\bibfnamefont {S.}~\bibnamefont
			{Yu}}, \bibinfo {author} {\bibfnamefont {S.}~\bibnamefont {H{\"{o}}fling}},
		\bibinfo {author} {\bibfnamefont {C.-Y.}\ \bibnamefont {Lu}},\ and\ \bibinfo
		{author} {\bibfnamefont {J.-W.}\ \bibnamefont {Pan}},\ }\bibfield  {title}
	{\bibinfo {title} {{Towards optimal single-photon sources from polarized
				microcavities}},\ }\href {https://doi.org/10.1038/s41566-019-0494-3}
	{\bibfield  {journal} {\bibinfo  {journal} {Nat. Photonics}\ }\textbf
		{\bibinfo {volume} {13}},\ \bibinfo {pages} {770} (\bibinfo {year}
		{2019})}\BibitemShut {NoStop}%
	\bibitem [{\citenamefont {Tiecke}\ \emph {et~al.}(2014)\citenamefont {Tiecke},
		\citenamefont {Thompson}, \citenamefont {de~Leon}, \citenamefont {Liu},
		\citenamefont {Vuleti{\'{c}}},\ and\ \citenamefont {Lukin}}]{Tiecke2014}%
	\BibitemOpen
	\bibfield  {author} {\bibinfo {author} {\bibfnamefont {T.~G.}\ \bibnamefont
			{Tiecke}}, \bibinfo {author} {\bibfnamefont {J.~D.}\ \bibnamefont
			{Thompson}}, \bibinfo {author} {\bibfnamefont {N.~P.}\ \bibnamefont
			{de~Leon}}, \bibinfo {author} {\bibfnamefont {L.~R.}\ \bibnamefont {Liu}},
		\bibinfo {author} {\bibfnamefont {V.}~\bibnamefont {Vuleti{\'{c}}}},\ and\
		\bibinfo {author} {\bibfnamefont {M.~D.}\ \bibnamefont {Lukin}},\ }\bibfield
	{title} {\bibinfo {title} {{Nanophotonic quantum phase switch with a single
				atom}},\ }\href {https://doi.org/10.1038/nature13188} {\bibfield  {journal}
		{\bibinfo  {journal} {Nature}\ }\textbf {\bibinfo {volume} {508}},\ \bibinfo
		{pages} {241} (\bibinfo {year} {2014})}\BibitemShut {NoStop}%
	\bibitem [{\citenamefont {Hummel}\ \emph {et~al.}(2019)\citenamefont {Hummel},
		\citenamefont {Ouellet-Plamondon}, \citenamefont {Ugur}, \citenamefont
		{Kulkova}, \citenamefont {Lund-Hansen}, \citenamefont {Broome}, \citenamefont
		{Uppu},\ and\ \citenamefont {Lodahl}}]{Hummel2019}%
	\BibitemOpen
	\bibfield  {author} {\bibinfo {author} {\bibfnamefont {T.}~\bibnamefont
			{Hummel}}, \bibinfo {author} {\bibfnamefont {C.}~\bibnamefont
			{Ouellet-Plamondon}}, \bibinfo {author} {\bibfnamefont {E.}~\bibnamefont
			{Ugur}}, \bibinfo {author} {\bibfnamefont {I.}~\bibnamefont {Kulkova}},
		\bibinfo {author} {\bibfnamefont {T.}~\bibnamefont {Lund-Hansen}}, \bibinfo
		{author} {\bibfnamefont {M.~A.}\ \bibnamefont {Broome}}, \bibinfo {author}
		{\bibfnamefont {R.}~\bibnamefont {Uppu}},\ and\ \bibinfo {author}
		{\bibfnamefont {P.}~\bibnamefont {Lodahl}},\ }\bibfield  {title} {\bibinfo
		{title} {{Efficient demultiplexed single-photon source with a quantum dot
				coupled to a nanophotonic waveguide}},\ }\href
	{https://doi.org/10.1063/1.5096979} {\bibfield  {journal} {\bibinfo
			{journal} {Appl. Phys. Lett.}\ }\textbf {\bibinfo {volume} {115}},\ \bibinfo
		{pages} {021102} (\bibinfo {year} {2019})}\BibitemShut {NoStop}%
	\bibitem [{\citenamefont {Bruzewicz}\ \emph {et~al.}(2019)\citenamefont
		{Bruzewicz}, \citenamefont {Chiaverini}, \citenamefont {McConnell},\ and\
		\citenamefont {Sage}}]{Bruzewicz2019}%
	\BibitemOpen
	\bibfield  {author} {\bibinfo {author} {\bibfnamefont {C.~D.}\ \bibnamefont
			{Bruzewicz}}, \bibinfo {author} {\bibfnamefont {J.}~\bibnamefont
			{Chiaverini}}, \bibinfo {author} {\bibfnamefont {R.}~\bibnamefont
			{McConnell}},\ and\ \bibinfo {author} {\bibfnamefont {J.~M.}\ \bibnamefont
			{Sage}},\ }\bibfield  {title} {\bibinfo {title} {{Trapped-ion quantum
				computing: Progress and challenges}},\ }\href
	{https://doi.org/10.1063/1.5088164} {\bibfield  {journal} {\bibinfo
			{journal} {App. Phys. Rev.}\ }\textbf {\bibinfo {volume} {6}},\ \bibinfo
		{pages} {021314} (\bibinfo {year} {2019})}\BibitemShut {NoStop}%
	\bibitem [{\citenamefont {Friis}\ \emph {et~al.}(2018)\citenamefont {Friis},
		\citenamefont {Marty}, \citenamefont {Maier}, \citenamefont {Hempel},
		\citenamefont {Holz{\"{a}}pfel}, \citenamefont {Jurcevic}, \citenamefont
		{Plenio}, \citenamefont {Huber}, \citenamefont {Roos}, \citenamefont
		{Blatt},\ and\ \citenamefont {Lanyon}}]{Friis2018}%
	\BibitemOpen
	\bibfield  {author} {\bibinfo {author} {\bibfnamefont {N.}~\bibnamefont
			{Friis}}, \bibinfo {author} {\bibfnamefont {O.}~\bibnamefont {Marty}},
		\bibinfo {author} {\bibfnamefont {C.}~\bibnamefont {Maier}}, \bibinfo
		{author} {\bibfnamefont {C.}~\bibnamefont {Hempel}}, \bibinfo {author}
		{\bibfnamefont {M.}~\bibnamefont {Holz{\"{a}}pfel}}, \bibinfo {author}
		{\bibfnamefont {P.}~\bibnamefont {Jurcevic}}, \bibinfo {author}
		{\bibfnamefont {M.~B.}\ \bibnamefont {Plenio}}, \bibinfo {author}
		{\bibfnamefont {M.}~\bibnamefont {Huber}}, \bibinfo {author} {\bibfnamefont
			{C.}~\bibnamefont {Roos}}, \bibinfo {author} {\bibfnamefont {R.}~\bibnamefont
			{Blatt}},\ and\ \bibinfo {author} {\bibfnamefont {B.}~\bibnamefont
			{Lanyon}},\ }\bibfield  {title} {\bibinfo {title} {{Observation of Entangled
				States of a Fully Controlled 20-Qubit System}},\ }\href
	{https://doi.org/10.1103/PhysRevX.8.021012} {\bibfield  {journal} {\bibinfo
			{journal} {Phys. Rev. X}\ }\textbf {\bibinfo {volume} {8}},\ \bibinfo {pages}
		{021012} (\bibinfo {year} {2018})}\BibitemShut {NoStop}%
	\bibitem [{\citenamefont {Zhang}\ \emph {et~al.}(2017)\citenamefont {Zhang},
		\citenamefont {Pagano}, \citenamefont {Hess}, \citenamefont {Kyprianidis},
		\citenamefont {Becker}, \citenamefont {Kaplan}, \citenamefont {Gorshkov},
		\citenamefont {Gong},\ and\ \citenamefont {Monroe}}]{Zhang2017}%
	\BibitemOpen
	\bibfield  {author} {\bibinfo {author} {\bibfnamefont {J.}~\bibnamefont
			{Zhang}}, \bibinfo {author} {\bibfnamefont {G.}~\bibnamefont {Pagano}},
		\bibinfo {author} {\bibfnamefont {P.~W.}\ \bibnamefont {Hess}}, \bibinfo
		{author} {\bibfnamefont {A.}~\bibnamefont {Kyprianidis}}, \bibinfo {author}
		{\bibfnamefont {P.}~\bibnamefont {Becker}}, \bibinfo {author} {\bibfnamefont
			{H.}~\bibnamefont {Kaplan}}, \bibinfo {author} {\bibfnamefont {A.~V.}\
			\bibnamefont {Gorshkov}}, \bibinfo {author} {\bibfnamefont {Z.-X.}\
			\bibnamefont {Gong}},\ and\ \bibinfo {author} {\bibfnamefont
			{C.}~\bibnamefont {Monroe}},\ }\bibfield  {title} {\bibinfo {title}
		{{Observation of a many-body dynamical phase transition with a 53-qubit
				quantum simulator}},\ }\href {https://doi.org/10.1038/nature24654} {\bibfield
		{journal} {\bibinfo  {journal} {Nature}\ }\textbf {\bibinfo {volume}
			{551}},\ \bibinfo {pages} {601} (\bibinfo {year} {2017})}\BibitemShut
	{NoStop}%
	\bibitem [{\citenamefont {H{\"{a}}ffner}\ \emph {et~al.}(2005)\citenamefont
		{H{\"{a}}ffner}, \citenamefont {Schmidt-Kaler}, \citenamefont {H{\"{a}}nsel},
		\citenamefont {Roos}, \citenamefont {K{\"{o}}rber}, \citenamefont {Chwalla},
		\citenamefont {Riebe}, \citenamefont {Benhelm}, \citenamefont {Rapol},
		\citenamefont {Becher},\ and\ \citenamefont {Blatt}}]{Haffner2005}%
	\BibitemOpen
	\bibfield  {author} {\bibinfo {author} {\bibfnamefont {H.}~\bibnamefont
			{H{\"{a}}ffner}}, \bibinfo {author} {\bibfnamefont {F.}~\bibnamefont
			{Schmidt-Kaler}}, \bibinfo {author} {\bibfnamefont {W.}~\bibnamefont
			{H{\"{a}}nsel}}, \bibinfo {author} {\bibfnamefont {C.~F.}\ \bibnamefont
			{Roos}}, \bibinfo {author} {\bibfnamefont {T.}~\bibnamefont {K{\"{o}}rber}},
		\bibinfo {author} {\bibfnamefont {M.}~\bibnamefont {Chwalla}}, \bibinfo
		{author} {\bibfnamefont {M.}~\bibnamefont {Riebe}}, \bibinfo {author}
		{\bibfnamefont {J.}~\bibnamefont {Benhelm}}, \bibinfo {author} {\bibfnamefont
			{U.~D.}\ \bibnamefont {Rapol}}, \bibinfo {author} {\bibfnamefont
			{C.}~\bibnamefont {Becher}},\ and\ \bibinfo {author} {\bibfnamefont
			{R.}~\bibnamefont {Blatt}},\ }\bibfield  {title} {\bibinfo {title} {{Robust
				entanglement}},\ }\href {https://doi.org/10.1007/s00340-005-1917-z}
	{\bibfield  {journal} {\bibinfo  {journal} {Appl. Phys. B}\ }\textbf
		{\bibinfo {volume} {81}},\ \bibinfo {pages} {151} (\bibinfo {year}
		{2005})}\BibitemShut {NoStop}%
	\bibitem [{\citenamefont {Sangouard}\ \emph {et~al.}(2009)\citenamefont
		{Sangouard}, \citenamefont {Dubessy},\ and\ \citenamefont
		{Simon}}]{Sangouard2009}%
	\BibitemOpen
	\bibfield  {author} {\bibinfo {author} {\bibfnamefont {N.}~\bibnamefont
			{Sangouard}}, \bibinfo {author} {\bibfnamefont {R.}~\bibnamefont {Dubessy}},\
		and\ \bibinfo {author} {\bibfnamefont {C.}~\bibnamefont {Simon}},\ }\bibfield
	{title} {\bibinfo {title} {{Quantum repeaters based on single trapped
				ions}},\ }\href {https://doi.org/10.1103/PhysRevA.79.042340} {\bibfield
		{journal} {\bibinfo  {journal} {Phys. Rev. A}\ }\textbf {\bibinfo {volume}
			{79}},\ \bibinfo {pages} {042340} (\bibinfo {year} {2009})}\BibitemShut
	{NoStop}%
	\bibitem [{\citenamefont {Munro}\ \emph {et~al.}(2015)\citenamefont {Munro},
		\citenamefont {Azuma}, \citenamefont {Tamaki},\ and\ \citenamefont
		{Nemoto}}]{Munro2015}%
	\BibitemOpen
	\bibfield  {author} {\bibinfo {author} {\bibfnamefont {W.~J.}\ \bibnamefont
			{Munro}}, \bibinfo {author} {\bibfnamefont {K.}~\bibnamefont {Azuma}},
		\bibinfo {author} {\bibfnamefont {K.}~\bibnamefont {Tamaki}},\ and\ \bibinfo
		{author} {\bibfnamefont {K.}~\bibnamefont {Nemoto}},\ }\bibfield  {title}
	{\bibinfo {title} {{Inside Quantum Repeaters}},\ }\href
	{https://doi.org/10.1109/JSTQE.2015.2392076} {\bibfield  {journal} {\bibinfo
			{journal} {IEEE J. Sel. Top. Quantum Electron.}\ }\textbf {\bibinfo {volume}
			{21}},\ \bibinfo {pages} {78} (\bibinfo {year} {2015})}\BibitemShut {NoStop}%
	\bibitem [{\citenamefont {Baumgart}\ \emph {et~al.}(2016)\citenamefont
		{Baumgart}, \citenamefont {Cai}, \citenamefont {Retzker}, \citenamefont
		{Plenio},\ and\ \citenamefont {Wunderlich}}]{Baumgart2016}%
	\BibitemOpen
	\bibfield  {author} {\bibinfo {author} {\bibfnamefont {I.}~\bibnamefont
			{Baumgart}}, \bibinfo {author} {\bibfnamefont {J.-M.}\ \bibnamefont {Cai}},
		\bibinfo {author} {\bibfnamefont {A.}~\bibnamefont {Retzker}}, \bibinfo
		{author} {\bibfnamefont {M.~B.}\ \bibnamefont {Plenio}},\ and\ \bibinfo
		{author} {\bibfnamefont {C.}~\bibnamefont {Wunderlich}},\ }\bibfield  {title}
	{\bibinfo {title} {{Ultrasensitive Magnetometer using a Single Atom}},\
	}\href {https://doi.org/10.1103/PhysRevLett.116.240801} {\bibfield  {journal}
		{\bibinfo  {journal} {Phys. Rev. Lett.}\ }\textbf {\bibinfo {volume} {116}},\
		\bibinfo {pages} {240801} (\bibinfo {year} {2016})}\BibitemShut {NoStop}%
	\bibitem [{\citenamefont {Brewer}\ \emph {et~al.}(2019)\citenamefont {Brewer},
		\citenamefont {Chen}, \citenamefont {Hankin}, \citenamefont {Clements},
		\citenamefont {Chou}, \citenamefont {Wineland}, \citenamefont {Hume},\ and\
		\citenamefont {Leibrandt}}]{Brewer2019}%
	\BibitemOpen
	\bibfield  {author} {\bibinfo {author} {\bibfnamefont {S.~M.}\ \bibnamefont
			{Brewer}}, \bibinfo {author} {\bibfnamefont {J.-S.}\ \bibnamefont {Chen}},
		\bibinfo {author} {\bibfnamefont {A.~M.}\ \bibnamefont {Hankin}}, \bibinfo
		{author} {\bibfnamefont {E.~R.}\ \bibnamefont {Clements}}, \bibinfo {author}
		{\bibfnamefont {C.~W.}\ \bibnamefont {Chou}}, \bibinfo {author}
		{\bibfnamefont {D.~J.}\ \bibnamefont {Wineland}}, \bibinfo {author}
		{\bibfnamefont {D.~B.}\ \bibnamefont {Hume}},\ and\ \bibinfo {author}
		{\bibfnamefont {D.~R.}\ \bibnamefont {Leibrandt}},\ }\bibfield  {title}
	{\bibinfo {title} {{27Al+ Quantum-Logic Clock with a Systematic Uncertainty
				below 10-18}},\ }\href {https://doi.org/10.1103/PhysRevLett.123.033201}
	{\bibfield  {journal} {\bibinfo  {journal} {Phys. Rev. Lett.}\ }\textbf
		{\bibinfo {volume} {123}},\ \bibinfo {pages} {033201} (\bibinfo {year}
		{2019})}\BibitemShut {NoStop}%
	\bibitem [{\citenamefont {Moehring}\ \emph {et~al.}(2007)\citenamefont
		{Moehring}, \citenamefont {Maunz}, \citenamefont {Olmschenk}, \citenamefont
		{Younge}, \citenamefont {Matsukevich}, \citenamefont {Duan},\ and\
		\citenamefont {Monroe}}]{Moehring2007}%
	\BibitemOpen
	\bibfield  {author} {\bibinfo {author} {\bibfnamefont {D.~L.}\ \bibnamefont
			{Moehring}}, \bibinfo {author} {\bibfnamefont {P.}~\bibnamefont {Maunz}},
		\bibinfo {author} {\bibfnamefont {S.}~\bibnamefont {Olmschenk}}, \bibinfo
		{author} {\bibfnamefont {K.~C.}\ \bibnamefont {Younge}}, \bibinfo {author}
		{\bibfnamefont {D.~N.}\ \bibnamefont {Matsukevich}}, \bibinfo {author}
		{\bibfnamefont {L.-M.}\ \bibnamefont {Duan}},\ and\ \bibinfo {author}
		{\bibfnamefont {C.}~\bibnamefont {Monroe}},\ }\bibfield  {title} {\bibinfo
		{title} {{Entanglement of single-atom quantum bits at a distance}},\ }\href
	{https://doi.org/10.1038/nature06118} {\bibfield  {journal} {\bibinfo
			{journal} {Nature}\ }\textbf {\bibinfo {volume} {449}},\ \bibinfo {pages}
		{68} (\bibinfo {year} {2007})}\BibitemShut {NoStop}%
	\bibitem [{\citenamefont {Hucul}\ \emph {et~al.}(2015)\citenamefont {Hucul},
		\citenamefont {Inlek}, \citenamefont {Vittorini}, \citenamefont {Crocker},
		\citenamefont {Debnath}, \citenamefont {Clark},\ and\ \citenamefont
		{Monroe}}]{Hucul2015}%
	\BibitemOpen
	\bibfield  {author} {\bibinfo {author} {\bibfnamefont {D.}~\bibnamefont
			{Hucul}}, \bibinfo {author} {\bibfnamefont {I.~V.}\ \bibnamefont {Inlek}},
		\bibinfo {author} {\bibfnamefont {G.}~\bibnamefont {Vittorini}}, \bibinfo
		{author} {\bibfnamefont {C.}~\bibnamefont {Crocker}}, \bibinfo {author}
		{\bibfnamefont {S.}~\bibnamefont {Debnath}}, \bibinfo {author} {\bibfnamefont
			{S.~M.}\ \bibnamefont {Clark}},\ and\ \bibinfo {author} {\bibfnamefont
			{C.}~\bibnamefont {Monroe}},\ }\bibfield  {title} {\bibinfo {title} {{Modular
				entanglement of atomic qubits using photons and phonons}},\ }\href
	{https://doi.org/10.1038/nphys3150} {\bibfield  {journal} {\bibinfo
			{journal} {Nat. Phys.}\ }\textbf {\bibinfo {volume} {11}},\ \bibinfo {pages}
		{37} (\bibinfo {year} {2015})}\BibitemShut {NoStop}%
	\bibitem [{\citenamefont {Stephenson}\ \emph {et~al.}(2020)\citenamefont
		{Stephenson}, \citenamefont {Nadlinger}, \citenamefont {Nichol},
		\citenamefont {An}, \citenamefont {Drmota}, \citenamefont {Ballance},
		\citenamefont {Thirumalai}, \citenamefont {Goodwin}, \citenamefont {Lucas},\
		and\ \citenamefont {Ballance}}]{Stephenson2020}%
	\BibitemOpen
	\bibfield  {author} {\bibinfo {author} {\bibfnamefont {L.~J.}\ \bibnamefont
			{Stephenson}}, \bibinfo {author} {\bibfnamefont {D.~P.}\ \bibnamefont
			{Nadlinger}}, \bibinfo {author} {\bibfnamefont {B.~C.}\ \bibnamefont
			{Nichol}}, \bibinfo {author} {\bibfnamefont {S.}~\bibnamefont {An}}, \bibinfo
		{author} {\bibfnamefont {P.}~\bibnamefont {Drmota}}, \bibinfo {author}
		{\bibfnamefont {T.~G.}\ \bibnamefont {Ballance}}, \bibinfo {author}
		{\bibfnamefont {K.}~\bibnamefont {Thirumalai}}, \bibinfo {author}
		{\bibfnamefont {J.~F.}\ \bibnamefont {Goodwin}}, \bibinfo {author}
		{\bibfnamefont {D.~M.}\ \bibnamefont {Lucas}},\ and\ \bibinfo {author}
		{\bibfnamefont {C.~J.}\ \bibnamefont {Ballance}},\ }\bibfield  {title}
	{\bibinfo {title} {{High-Rate, High-Fidelity Entanglement of Qubits Across an
				Elementary Quantum Network}},\ }\href
	{https://doi.org/10.1103/PhysRevLett.124.110501} {\bibfield  {journal}
		{\bibinfo  {journal} {Phys. Rev. Lett.}\ }\textbf {\bibinfo {volume} {124}},\
		\bibinfo {pages} {110501} (\bibinfo {year} {2020})}\BibitemShut {NoStop}%
	\bibitem [{\citenamefont {Stute}\ \emph {et~al.}(2012)\citenamefont {Stute},
		\citenamefont {Casabone}, \citenamefont {Schindler}, \citenamefont {Monz},
		\citenamefont {Schmidt}, \citenamefont {Brandst{\"{a}}tter}, \citenamefont
		{Northup},\ and\ \citenamefont {Blatt}}]{Stute2012}%
	\BibitemOpen
	\bibfield  {author} {\bibinfo {author} {\bibfnamefont {A.}~\bibnamefont
			{Stute}}, \bibinfo {author} {\bibfnamefont {B.}~\bibnamefont {Casabone}},
		\bibinfo {author} {\bibfnamefont {P.}~\bibnamefont {Schindler}}, \bibinfo
		{author} {\bibfnamefont {T.}~\bibnamefont {Monz}}, \bibinfo {author}
		{\bibfnamefont {P.~O.}\ \bibnamefont {Schmidt}}, \bibinfo {author}
		{\bibfnamefont {B.}~\bibnamefont {Brandst{\"{a}}tter}}, \bibinfo {author}
		{\bibfnamefont {T.~E.}\ \bibnamefont {Northup}},\ and\ \bibinfo {author}
		{\bibfnamefont {R.}~\bibnamefont {Blatt}},\ }\bibfield  {title} {\bibinfo
		{title} {{Tunable ion–photon entanglement in an optical cavity}},\ }\href
	{https://doi.org/10.1038/nature11120} {\bibfield  {journal} {\bibinfo
			{journal} {Nature}\ }\textbf {\bibinfo {volume} {485}},\ \bibinfo {pages}
		{482} (\bibinfo {year} {2012})}\BibitemShut {NoStop}%
	\bibitem [{\citenamefont {Krutyanskiy}\ \emph {et~al.}(2019)\citenamefont
		{Krutyanskiy}, \citenamefont {Meraner}, \citenamefont {Schupp}, \citenamefont
		{Krcmarsky}, \citenamefont {Hainzer},\ and\ \citenamefont
		{Lanyon}}]{Krutyanskiy2019}%
	\BibitemOpen
	\bibfield  {author} {\bibinfo {author} {\bibfnamefont {V.}~\bibnamefont
			{Krutyanskiy}}, \bibinfo {author} {\bibfnamefont {M.}~\bibnamefont
			{Meraner}}, \bibinfo {author} {\bibfnamefont {J.}~\bibnamefont {Schupp}},
		\bibinfo {author} {\bibfnamefont {V.}~\bibnamefont {Krcmarsky}}, \bibinfo
		{author} {\bibfnamefont {H.}~\bibnamefont {Hainzer}},\ and\ \bibinfo {author}
		{\bibfnamefont {B.~P.}\ \bibnamefont {Lanyon}},\ }\bibfield  {title}
	{\bibinfo {title} {{Light-matter entanglement over 50 km of optical fibre}},\
	}\href {https://doi.org/10.1038/s41534-019-0186-3} {\bibfield  {journal}
		{\bibinfo  {journal} {npj Quantum Inf.}\ }\textbf {\bibinfo {volume} {5}},\
		\bibinfo {pages} {72} (\bibinfo {year} {2019})}\BibitemShut {NoStop}%
	\bibitem [{eff()}]{efficiency50km}%
	\BibitemOpen
	\href@noop {} {}\bibinfo {note} {The value 0.08(1) is calculated from the
		detected efficiency at 1550 nm after 50 km and using the efficiencies in
		TABLE I in the supplementary material of \cite{Krutyanskiy2019} and a
		detector efficiency of 0.4.}\BibitemShut {Stop}%
	\bibitem [{\citenamefont {Kobel}\ \emph {et~al.}(2021)\citenamefont {Kobel},
		\citenamefont {Breyer},\ and\ \citenamefont {K{\"{o}}hl}}]{Kobel2021}%
	\BibitemOpen
	\bibfield  {author} {\bibinfo {author} {\bibfnamefont {P.}~\bibnamefont
			{Kobel}}, \bibinfo {author} {\bibfnamefont {M.}~\bibnamefont {Breyer}},\ and\
		\bibinfo {author} {\bibfnamefont {M.}~\bibnamefont {K{\"{o}}hl}},\ }\bibfield
	{title} {\bibinfo {title} {{Deterministic spin-photon entanglement from a
				trapped ion in a fiber Fabry–Perot cavity}},\ }\href
	{https://doi.org/10.1038/s41534-020-00338-2} {\bibfield  {journal} {\bibinfo
			{journal} {npj Quantum Inf.}\ }\textbf {\bibinfo {volume} {7}},\ \bibinfo
		{pages} {6} (\bibinfo {year} {2021})}\BibitemShut {NoStop}%
	\bibitem [{\citenamefont {Goto}\ \emph {et~al.}(2019)\citenamefont {Goto},
		\citenamefont {Mizukami}, \citenamefont {Tokunaga},\ and\ \citenamefont
		{Aoki}}]{Goto2019}%
	\BibitemOpen
	\bibfield  {author} {\bibinfo {author} {\bibfnamefont {H.}~\bibnamefont
			{Goto}}, \bibinfo {author} {\bibfnamefont {S.}~\bibnamefont {Mizukami}},
		\bibinfo {author} {\bibfnamefont {Y.}~\bibnamefont {Tokunaga}},\ and\
		\bibinfo {author} {\bibfnamefont {T.}~\bibnamefont {Aoki}},\ }\bibfield
	{title} {\bibinfo {title} {{Figure of merit for single-photon generation
				based on cavity quantum electrodynamics}},\ }\href
	{https://doi.org/10.1103/PhysRevA.99.053843} {\bibfield  {journal} {\bibinfo
			{journal} {Phys. Rev. A}\ }\textbf {\bibinfo {volume} {99}},\ \bibinfo
		{pages} {053843} (\bibinfo {year} {2019})}\BibitemShut {NoStop}%
	\bibitem [{\citenamefont {Law}\ and\ \citenamefont {Kimble}(1997)}]{Law1997}%
	\BibitemOpen
	\bibfield  {author} {\bibinfo {author} {\bibfnamefont {C.~K.}\ \bibnamefont
			{Law}}\ and\ \bibinfo {author} {\bibfnamefont {H.~J.}\ \bibnamefont
			{Kimble}},\ }\bibfield  {title} {\bibinfo {title} {{Deterministic generation
				of a bit-stream of single-photon pulses}},\ }\href
	{https://doi.org/10.1080/09500349708231869} {\bibfield  {journal} {\bibinfo
			{journal} {J. Mod. Opt.}\ }\textbf {\bibinfo {volume} {44}},\ \bibinfo
		{pages} {2067} (\bibinfo {year} {1997})}\BibitemShut {NoStop}%
	\bibitem [{\citenamefont {Kuhn}\ and\ \citenamefont
		{Ljunggren}(2010)}]{Kuhn2010}%
	\BibitemOpen
	\bibfield  {author} {\bibinfo {author} {\bibfnamefont {A.}~\bibnamefont
			{Kuhn}}\ and\ \bibinfo {author} {\bibfnamefont {D.}~\bibnamefont
			{Ljunggren}},\ }\bibfield  {title} {\bibinfo {title} {{Cavity-based
				single-photon sources}},\ }\href {https://doi.org/10.1080/00107511003602990}
	{\bibfield  {journal} {\bibinfo  {journal} {Contemp. Phys.}\ }\textbf
		{\bibinfo {volume} {51}},\ \bibinfo {pages} {289} (\bibinfo {year}
		{2010})}\BibitemShut {NoStop}%
	\bibitem [{\citenamefont {Vasilev}\ \emph {et~al.}(2010)\citenamefont
		{Vasilev}, \citenamefont {Ljunggren},\ and\ \citenamefont
		{Kuhn}}]{Vasilev2010}%
	\BibitemOpen
	\bibfield  {author} {\bibinfo {author} {\bibfnamefont {G.~S.}\ \bibnamefont
			{Vasilev}}, \bibinfo {author} {\bibfnamefont {D.}~\bibnamefont {Ljunggren}},\
		and\ \bibinfo {author} {\bibfnamefont {A.}~\bibnamefont {Kuhn}},\ }\bibfield
	{title} {\bibinfo {title} {{Single photons made-to-measure}},\ }\href
	{https://doi.org/10.1088/1367-2630/12/6/063024} {\bibfield  {journal}
		{\bibinfo  {journal} {New J. Phys.}\ }\textbf {\bibinfo {volume} {12}},\
		\bibinfo {pages} {063024} (\bibinfo {year} {2010})}\BibitemShut {NoStop}%
	\bibitem [{\citenamefont {Meraner}\ \emph {et~al.}(2020)\citenamefont
		{Meraner}, \citenamefont {Mazloom}, \citenamefont {Krutyanskiy},
		\citenamefont {Krcmarsky}, \citenamefont {Schupp}, \citenamefont {Fioretto},
		\citenamefont {Sekatski}, \citenamefont {Northup}, \citenamefont
		{Sangouard},\ and\ \citenamefont {Lanyon}}]{Meraner2020}%
	\BibitemOpen
	\bibfield  {author} {\bibinfo {author} {\bibfnamefont {M.}~\bibnamefont
			{Meraner}}, \bibinfo {author} {\bibfnamefont {A.}~\bibnamefont {Mazloom}},
		\bibinfo {author} {\bibfnamefont {V.}~\bibnamefont {Krutyanskiy}}, \bibinfo
		{author} {\bibfnamefont {V.}~\bibnamefont {Krcmarsky}}, \bibinfo {author}
		{\bibfnamefont {J.}~\bibnamefont {Schupp}}, \bibinfo {author} {\bibfnamefont
			{D.~A.}\ \bibnamefont {Fioretto}}, \bibinfo {author} {\bibfnamefont
			{P.}~\bibnamefont {Sekatski}}, \bibinfo {author} {\bibfnamefont {T.~E.}\
			\bibnamefont {Northup}}, \bibinfo {author} {\bibfnamefont {N.}~\bibnamefont
			{Sangouard}},\ and\ \bibinfo {author} {\bibfnamefont {B.~P.}\ \bibnamefont
			{Lanyon}},\ }\bibfield  {title} {\bibinfo {title} {{Indistinguishable photons
				from a trapped-ion quantum network node}},\ }\href
	{https://doi.org/10.1103/PhysRevA.102.052614} {\bibfield  {journal} {\bibinfo
			{journal} {Phys. Rev. A}\ }\textbf {\bibinfo {volume} {102}},\ \bibinfo
		{pages} {052614} (\bibinfo {year} {2020})}\BibitemShut {NoStop}%
	\bibitem [{\citenamefont {Walker}\ \emph {et~al.}(2018)\citenamefont {Walker},
		\citenamefont {Miyanishi}, \citenamefont {Ikuta}, \citenamefont {Takahashi},
		\citenamefont {{Vartabi Kashanian}}, \citenamefont {Tsujimoto}, \citenamefont
		{Hayasaka}, \citenamefont {Yamamoto}, \citenamefont {Imoto},\ and\
		\citenamefont {Keller}}]{Walker2018}%
	\BibitemOpen
	\bibfield  {author} {\bibinfo {author} {\bibfnamefont {T.}~\bibnamefont
			{Walker}}, \bibinfo {author} {\bibfnamefont {K.}~\bibnamefont {Miyanishi}},
		\bibinfo {author} {\bibfnamefont {R.}~\bibnamefont {Ikuta}}, \bibinfo
		{author} {\bibfnamefont {H.}~\bibnamefont {Takahashi}}, \bibinfo {author}
		{\bibfnamefont {S.}~\bibnamefont {{Vartabi Kashanian}}}, \bibinfo {author}
		{\bibfnamefont {Y.}~\bibnamefont {Tsujimoto}}, \bibinfo {author}
		{\bibfnamefont {K.}~\bibnamefont {Hayasaka}}, \bibinfo {author}
		{\bibfnamefont {T.}~\bibnamefont {Yamamoto}}, \bibinfo {author}
		{\bibfnamefont {N.}~\bibnamefont {Imoto}},\ and\ \bibinfo {author}
		{\bibfnamefont {M.}~\bibnamefont {Keller}},\ }\bibfield  {title} {\bibinfo
		{title} {{Long-Distance Single Photon Transmission from a Trapped Ion via
				Quantum Frequency Conversion}},\ }\href
	{https://doi.org/10.1103/PhysRevLett.120.203601} {\bibfield  {journal}
		{\bibinfo  {journal} {Phys. Rev. Lett.}\ }\textbf {\bibinfo {volume} {120}},\
		\bibinfo {pages} {203601} (\bibinfo {year} {2018})}\BibitemShut {NoStop}%
	\bibitem [{\citenamefont {Bock}\ \emph {et~al.}(2018)\citenamefont {Bock},
		\citenamefont {Eich}, \citenamefont {Kucera}, \citenamefont {Kreis},
		\citenamefont {Lenhard}, \citenamefont {Becher},\ and\ \citenamefont
		{Eschner}}]{Bock2018}%
	\BibitemOpen
	\bibfield  {author} {\bibinfo {author} {\bibfnamefont {M.}~\bibnamefont
			{Bock}}, \bibinfo {author} {\bibfnamefont {P.}~\bibnamefont {Eich}}, \bibinfo
		{author} {\bibfnamefont {S.}~\bibnamefont {Kucera}}, \bibinfo {author}
		{\bibfnamefont {M.}~\bibnamefont {Kreis}}, \bibinfo {author} {\bibfnamefont
			{A.}~\bibnamefont {Lenhard}}, \bibinfo {author} {\bibfnamefont
			{C.}~\bibnamefont {Becher}},\ and\ \bibinfo {author} {\bibfnamefont
			{J.}~\bibnamefont {Eschner}},\ }\bibfield  {title} {\bibinfo {title}
		{{High-fidelity entanglement between a trapped ion and a telecom photon via
				quantum frequency conversion}},\ }\href
	{https://doi.org/10.1038/s41467-018-04341-2} {\bibfield  {journal} {\bibinfo
			{journal} {Nat. Commun.}\ }\textbf {\bibinfo {volume} {9}},\ \bibinfo {pages}
		{1998} (\bibinfo {year} {2018})}\BibitemShut {NoStop}%
	\bibitem [{\citenamefont {Keller}\ \emph {et~al.}(2004)\citenamefont {Keller},
		\citenamefont {Lange}, \citenamefont {Hayasaka}, \citenamefont {Lange},\ and\
		\citenamefont {Walther}}]{Keller2004}%
	\BibitemOpen
	\bibfield  {author} {\bibinfo {author} {\bibfnamefont {M.}~\bibnamefont
			{Keller}}, \bibinfo {author} {\bibfnamefont {B.}~\bibnamefont {Lange}},
		\bibinfo {author} {\bibfnamefont {K.}~\bibnamefont {Hayasaka}}, \bibinfo
		{author} {\bibfnamefont {W.}~\bibnamefont {Lange}},\ and\ \bibinfo {author}
		{\bibfnamefont {H.}~\bibnamefont {Walther}},\ }\bibfield  {title} {\bibinfo
		{title} {{Continuous generation of single photons with controlled waveform in
				an ion-trap cavity system}},\ }\href {https://doi.org/10.1038/nature02961}
	{\bibfield  {journal} {\bibinfo  {journal} {Nature}\ }\textbf {\bibinfo
			{volume} {431}},\ \bibinfo {pages} {1075} (\bibinfo {year}
		{2004})}\BibitemShut {NoStop}%
	\bibitem [{\citenamefont {Barros}\ \emph {et~al.}(2009)\citenamefont {Barros},
		\citenamefont {Stute}, \citenamefont {Northup}, \citenamefont {Russo},
		\citenamefont {Schmidt},\ and\ \citenamefont {Blatt}}]{Barros2009}%
	\BibitemOpen
	\bibfield  {author} {\bibinfo {author} {\bibfnamefont {H.~G.}\ \bibnamefont
			{Barros}}, \bibinfo {author} {\bibfnamefont {A.}~\bibnamefont {Stute}},
		\bibinfo {author} {\bibfnamefont {T.~E.}\ \bibnamefont {Northup}}, \bibinfo
		{author} {\bibfnamefont {C.}~\bibnamefont {Russo}}, \bibinfo {author}
		{\bibfnamefont {P.~O.}\ \bibnamefont {Schmidt}},\ and\ \bibinfo {author}
		{\bibfnamefont {R.}~\bibnamefont {Blatt}},\ }\bibfield  {title} {\bibinfo
		{title} {{Deterministic single-photon source from a single ion}},\ }\href
	{https://doi.org/10.1088/1367-2630/11/10/103004} {\bibfield  {journal}
		{\bibinfo  {journal} {New J. Phys.}\ }\textbf {\bibinfo {volume} {11}},\
		\bibinfo {pages} {103004} (\bibinfo {year} {2009})}\BibitemShut {NoStop}%
	\bibitem [{\citenamefont {Hainzer}(2018)}]{Hainzer2018}%
	\BibitemOpen
	\bibfield  {author} {\bibinfo {author} {\bibfnamefont {H.}~\bibnamefont
			{Hainzer}},\ }\emph {\bibinfo {title} {{Laser Locking For Trapped-Ion Quantum
				Networks}}},\ \href@noop {} {\bibinfo {type} {Master's thesis}},\ \bibinfo
	{school} {University of Innsbruck} (\bibinfo {year} {2018})\BibitemShut
	{NoStop}%
	\bibitem [{\citenamefont {Jin}\ and\ \citenamefont {Church}(1993)}]{Jin1993}%
	\BibitemOpen
	\bibfield  {author} {\bibinfo {author} {\bibfnamefont {J.}~\bibnamefont
			{Jin}}\ and\ \bibinfo {author} {\bibfnamefont {D.~A.}\ \bibnamefont
			{Church}},\ }\bibfield  {title} {\bibinfo {title} {{Precision Lifetimes for
				the Ca+ 4p 2P Levels: Experiment Challenges Theory at the 1{\%} Level}},\
	}\href {https://doi.org/10.1103/PhysRevLett.70.3213} {\bibfield  {journal}
		{\bibinfo  {journal} {Phys. Rev. Lett.}\ }\textbf {\bibinfo {volume} {70}},\
		\bibinfo {pages} {3213} (\bibinfo {year} {1993})}\BibitemShut {NoStop}%
	\bibitem [{\citenamefont {Brandst{\"{a}}tter}(2013)}]{Brandstatter2013}%
	\BibitemOpen
	\bibfield  {author} {\bibinfo {author} {\bibfnamefont {B.~U.}\ \bibnamefont
			{Brandst{\"{a}}tter}},\ }\emph {\bibinfo {title} {{Integration of fiber
				mirrors and ion traps for a high-fidelity quantum interface}}},\ \href@noop
	{} {\bibinfo {type} {Phd}},\ \bibinfo  {school} {Univeristy of Innsbruck}
	(\bibinfo {year} {2013})\BibitemShut {NoStop}%
	\bibitem [{\citenamefont {Nisbet-Jones}\ \emph {et~al.}(2011)\citenamefont
		{Nisbet-Jones}, \citenamefont {Dilley}, \citenamefont {Ljunggren},\ and\
		\citenamefont {Kuhn}}]{Nisbet-Jones2011}%
	\BibitemOpen
	\bibfield  {author} {\bibinfo {author} {\bibfnamefont {P.~B.~R.}\
			\bibnamefont {Nisbet-Jones}}, \bibinfo {author} {\bibfnamefont
			{J.}~\bibnamefont {Dilley}}, \bibinfo {author} {\bibfnamefont
			{D.}~\bibnamefont {Ljunggren}},\ and\ \bibinfo {author} {\bibfnamefont
			{A.}~\bibnamefont {Kuhn}},\ }\bibfield  {title} {\bibinfo {title} {{Highly
				efficient source for indistinguishable single photons of controlled shape}},\
	}\href {https://doi.org/10.1088/1367-2630/13/10/103036} {\bibfield  {journal}
		{\bibinfo  {journal} {New J. Phys.}\ }\textbf {\bibinfo {volume} {13}},\
		\bibinfo {pages} {103036} (\bibinfo {year} {2011})}\BibitemShut {NoStop}%
	\bibitem [{\citenamefont {Morin}\ \emph {et~al.}(2019)\citenamefont {Morin},
		\citenamefont {K{\"{o}}rber}, \citenamefont {Langenfeld},\ and\ \citenamefont
		{Rempe}}]{Morin2019}%
	\BibitemOpen
	\bibfield  {author} {\bibinfo {author} {\bibfnamefont {O.}~\bibnamefont
			{Morin}}, \bibinfo {author} {\bibfnamefont {M.}~\bibnamefont {K{\"{o}}rber}},
		\bibinfo {author} {\bibfnamefont {S.}~\bibnamefont {Langenfeld}},\ and\
		\bibinfo {author} {\bibfnamefont {G.}~\bibnamefont {Rempe}},\ }\bibfield
	{title} {\bibinfo {title} {{Deterministic Shaping and Reshaping of
				Single-Photon Temporal Wave Functions}},\ }\href
	{https://doi.org/10.1103/PhysRevLett.123.133602} {\bibfield  {journal}
		{\bibinfo  {journal} {Phys. Rev. Lett.}\ }\textbf {\bibinfo {volume} {123}},\
		\bibinfo {pages} {133602} (\bibinfo {year} {2019})}\BibitemShut {NoStop}%
	\bibitem [{\citenamefont {Schindler}\ \emph {et~al.}(2013)\citenamefont
		{Schindler}, \citenamefont {Nigg}, \citenamefont {Monz}, \citenamefont
		{Barreiro}, \citenamefont {Martinez}, \citenamefont {Wang}, \citenamefont
		{Quint}, \citenamefont {Brandl}, \citenamefont {Nebendahl}, \citenamefont
		{Roos}, \citenamefont {Chwalla}, \citenamefont {Hennrich},\ and\
		\citenamefont {Blatt}}]{Schindler2013}%
	\BibitemOpen
	\bibfield  {author} {\bibinfo {author} {\bibfnamefont {P.}~\bibnamefont
			{Schindler}}, \bibinfo {author} {\bibfnamefont {D.}~\bibnamefont {Nigg}},
		\bibinfo {author} {\bibfnamefont {T.}~\bibnamefont {Monz}}, \bibinfo {author}
		{\bibfnamefont {J.~T.}\ \bibnamefont {Barreiro}}, \bibinfo {author}
		{\bibfnamefont {E.}~\bibnamefont {Martinez}}, \bibinfo {author}
		{\bibfnamefont {S.~X.}\ \bibnamefont {Wang}}, \bibinfo {author}
		{\bibfnamefont {S.}~\bibnamefont {Quint}}, \bibinfo {author} {\bibfnamefont
			{M.~F.}\ \bibnamefont {Brandl}}, \bibinfo {author} {\bibfnamefont
			{V.}~\bibnamefont {Nebendahl}}, \bibinfo {author} {\bibfnamefont {C.~F.}\
			\bibnamefont {Roos}}, \bibinfo {author} {\bibfnamefont {M.}~\bibnamefont
			{Chwalla}}, \bibinfo {author} {\bibfnamefont {M.}~\bibnamefont {Hennrich}},\
		and\ \bibinfo {author} {\bibfnamefont {R.}~\bibnamefont {Blatt}},\ }\bibfield
	{title} {\bibinfo {title} {{A quantum information processor with trapped
				ions}},\ }\href {https://doi.org/10.1088/1367-2630/15/12/123012} {\bibfield
		{journal} {\bibinfo  {journal} {New J. Phys.}\ }\textbf {\bibinfo {volume}
			{15}},\ \bibinfo {pages} {123012} (\bibinfo {year} {2013})}\BibitemShut
	{NoStop}%
	\bibitem [{\citenamefont {Efron}\ and\ \citenamefont
		{Tibshirani}(1986)}]{Efron1986}%
	\BibitemOpen
	\bibfield  {author} {\bibinfo {author} {\bibfnamefont {B.}~\bibnamefont
			{Efron}}\ and\ \bibinfo {author} {\bibfnamefont {R.}~\bibnamefont
			{Tibshirani}},\ }\bibfield  {title} {\bibinfo {title} {{Bootstrap Methods for
				Standard Errors, Confidence Intervals, and Other Measures of Statistical
				Accuracy}},\ }\href {https://doi.org/10.1214/ss/1177013815} {\bibfield
		{journal} {\bibinfo  {journal} {Statistical Science}\ }\textbf {\bibinfo
			{volume} {1}},\ \bibinfo {pages} {54} (\bibinfo {year} {1986})}\BibitemShut
	{NoStop}%
	\bibitem [{\citenamefont {Lindner}\ and\ \citenamefont
		{Rudolph}(2009)}]{Lindner2009}%
	\BibitemOpen
	\bibfield  {author} {\bibinfo {author} {\bibfnamefont {N.~H.}\ \bibnamefont
			{Lindner}}\ and\ \bibinfo {author} {\bibfnamefont {T.}~\bibnamefont
			{Rudolph}},\ }\bibfield  {title} {\bibinfo {title} {{Proposal for Pulsed
				On-Demand Sources of Photonic Cluster State Strings}},\ }\href
	{https://doi.org/10.1103/PhysRevLett.103.113602} {\bibfield  {journal}
		{\bibinfo  {journal} {Phys. Rev. Lett.}\ }\textbf {\bibinfo {volume} {103}},\
		\bibinfo {pages} {113602} (\bibinfo {year} {2009})}\BibitemShut {NoStop}%
	\bibitem [{\citenamefont {Economou}\ \emph {et~al.}(2010)\citenamefont
		{Economou}, \citenamefont {Lindner},\ and\ \citenamefont
		{Rudolph}}]{Economou2010}%
	\BibitemOpen
	\bibfield  {author} {\bibinfo {author} {\bibfnamefont {S.~E.}\ \bibnamefont
			{Economou}}, \bibinfo {author} {\bibfnamefont {N.}~\bibnamefont {Lindner}},\
		and\ \bibinfo {author} {\bibfnamefont {T.}~\bibnamefont {Rudolph}},\
	}\bibfield  {title} {\bibinfo {title} {{Optically Generated 2-Dimensional
				Photonic Cluster State from Coupled Quantum Dots}},\ }\href
	{https://doi.org/10.1103/PhysRevLett.105.093601} {\bibfield  {journal}
		{\bibinfo  {journal} {Phys. Rev. Lett.}\ }\textbf {\bibinfo {volume} {105}},\
		\bibinfo {pages} {093601} (\bibinfo {year} {2010})}\BibitemShut {NoStop}%
	\bibitem [{\citenamefont {Schwartz}\ \emph {et~al.}(2016)\citenamefont
		{Schwartz}, \citenamefont {Cogan}, \citenamefont {Schmidgall}, \citenamefont
		{Don}, \citenamefont {Gantz}, \citenamefont {Kenneth}, \citenamefont
		{Lindner},\ and\ \citenamefont {Gershoni}}]{Schwartz2016}%
	\BibitemOpen
	\bibfield  {author} {\bibinfo {author} {\bibfnamefont {I.}~\bibnamefont
			{Schwartz}}, \bibinfo {author} {\bibfnamefont {D.}~\bibnamefont {Cogan}},
		\bibinfo {author} {\bibfnamefont {E.~R.}\ \bibnamefont {Schmidgall}},
		\bibinfo {author} {\bibfnamefont {Y.}~\bibnamefont {Don}}, \bibinfo {author}
		{\bibfnamefont {L.}~\bibnamefont {Gantz}}, \bibinfo {author} {\bibfnamefont
			{O.}~\bibnamefont {Kenneth}}, \bibinfo {author} {\bibfnamefont {N.~H.}\
			\bibnamefont {Lindner}},\ and\ \bibinfo {author} {\bibfnamefont
			{D.}~\bibnamefont {Gershoni}},\ }\bibfield  {title} {\bibinfo {title}
		{{Deterministic generation of a cluster state of entangled photons}},\ }\href
	{https://doi.org/10.1126/science.aah4758} {\bibfield  {journal} {\bibinfo
			{journal} {Science}\ }\textbf {\bibinfo {volume} {354}},\ \bibinfo {pages}
		{434} (\bibinfo {year} {2016})}\BibitemShut {NoStop}%
	\bibitem [{\citenamefont {Azuma}\ \emph {et~al.}(2015)\citenamefont {Azuma},
		\citenamefont {Tamaki},\ and\ \citenamefont {Lo}}]{Azuma2015}%
	\BibitemOpen
	\bibfield  {author} {\bibinfo {author} {\bibfnamefont {K.}~\bibnamefont
			{Azuma}}, \bibinfo {author} {\bibfnamefont {K.}~\bibnamefont {Tamaki}},\ and\
		\bibinfo {author} {\bibfnamefont {H.-k.}\ \bibnamefont {Lo}},\ }\bibfield
	{title} {\bibinfo {title} {{All-photonic quantum repeaters}},\ }\href
	{https://doi.org/10.1038/ncomms7787} {\bibfield  {journal} {\bibinfo
			{journal} {Nat. Commun.}\ }\textbf {\bibinfo {volume} {6}},\ \bibinfo {pages}
		{6787} (\bibinfo {year} {2015})}\BibitemShut {NoStop}%
	\bibitem [{\citenamefont {Borregaard}\ \emph {et~al.}(2020)\citenamefont
		{Borregaard}, \citenamefont {Pichler}, \citenamefont {Schr{\"{o}}der},
		\citenamefont {Lukin}, \citenamefont {Lodahl},\ and\ \citenamefont
		{S{\o}rensen}}]{Borregaard2020}%
	\BibitemOpen
	\bibfield  {author} {\bibinfo {author} {\bibfnamefont {J.}~\bibnamefont
			{Borregaard}}, \bibinfo {author} {\bibfnamefont {H.}~\bibnamefont {Pichler}},
		\bibinfo {author} {\bibfnamefont {T.}~\bibnamefont {Schr{\"{o}}der}},
		\bibinfo {author} {\bibfnamefont {M.~D.}\ \bibnamefont {Lukin}}, \bibinfo
		{author} {\bibfnamefont {P.}~\bibnamefont {Lodahl}},\ and\ \bibinfo {author}
		{\bibfnamefont {A.~S.}\ \bibnamefont {S{\o}rensen}},\ }\bibfield  {title}
	{\bibinfo {title} {{One-Way Quantum Repeater Based on Near-Deterministic
				Photon-Emitter Interfaces}},\ }\href
	{https://doi.org/10.1103/PhysRevX.10.021071} {\bibfield  {journal} {\bibinfo
			{journal} {Phys. Rev. X}\ }\textbf {\bibinfo {volume} {10}},\ \bibinfo
		{pages} {021071} (\bibinfo {year} {2020})}\BibitemShut {NoStop}%
	\bibitem [{dur()}]{duration}%
	\BibitemOpen
	\href@noop {} {}\bibinfo {note} {During $\sim50\%$ of the experiment
		duration, no photon generation attempts were made due to the reprogramming
		time of our current control system.}\BibitemShut {Stop}%
	\bibitem [{\citenamefont {Lechner}\ \emph {et~al.}(2016)\citenamefont
		{Lechner}, \citenamefont {Maier}, \citenamefont {Hempel}, \citenamefont
		{Jurcevic}, \citenamefont {Lanyon}, \citenamefont {Monz}, \citenamefont
		{Brownnutt}, \citenamefont {Blatt},\ and\ \citenamefont
		{Roos}}]{Lechner2016}%
	\BibitemOpen
	\bibfield  {author} {\bibinfo {author} {\bibfnamefont {R.}~\bibnamefont
			{Lechner}}, \bibinfo {author} {\bibfnamefont {C.}~\bibnamefont {Maier}},
		\bibinfo {author} {\bibfnamefont {C.}~\bibnamefont {Hempel}}, \bibinfo
		{author} {\bibfnamefont {P.}~\bibnamefont {Jurcevic}}, \bibinfo {author}
		{\bibfnamefont {B.~P.}\ \bibnamefont {Lanyon}}, \bibinfo {author}
		{\bibfnamefont {T.}~\bibnamefont {Monz}}, \bibinfo {author} {\bibfnamefont
			{M.}~\bibnamefont {Brownnutt}}, \bibinfo {author} {\bibfnamefont
			{R.}~\bibnamefont {Blatt}},\ and\ \bibinfo {author} {\bibfnamefont {C.~F.}\
			\bibnamefont {Roos}},\ }\bibfield  {title} {\bibinfo {title}
		{{Electromagnetically-induced-transparency ground-state cooling of long ion
				strings}},\ }\href {https://doi.org/10.1103/PhysRevA.93.053401} {\bibfield
		{journal} {\bibinfo  {journal} {Phys. Rev. A}\ }\textbf {\bibinfo {volume}
			{93}},\ \bibinfo {pages} {053401} (\bibinfo {year} {2016})}\BibitemShut
	{NoStop}%
	\bibitem [{\citenamefont {Joshi}\ \emph {et~al.}(2020)\citenamefont {Joshi},
		\citenamefont {Fabre}, \citenamefont {Maier}, \citenamefont {Brydges},
		\citenamefont {Kiesenhofer}, \citenamefont {Hainzer}, \citenamefont {Blatt},\
		and\ \citenamefont {Roos}}]{Joshi2020}%
	\BibitemOpen
	\bibfield  {author} {\bibinfo {author} {\bibfnamefont {M.~K.}\ \bibnamefont
			{Joshi}}, \bibinfo {author} {\bibfnamefont {A.}~\bibnamefont {Fabre}},
		\bibinfo {author} {\bibfnamefont {C.}~\bibnamefont {Maier}}, \bibinfo
		{author} {\bibfnamefont {T.}~\bibnamefont {Brydges}}, \bibinfo {author}
		{\bibfnamefont {D.}~\bibnamefont {Kiesenhofer}}, \bibinfo {author}
		{\bibfnamefont {H.}~\bibnamefont {Hainzer}}, \bibinfo {author} {\bibfnamefont
			{R.}~\bibnamefont {Blatt}},\ and\ \bibinfo {author} {\bibfnamefont {C.~F.}\
			\bibnamefont {Roos}},\ }\bibfield  {title} {\bibinfo {title}
		{{Polarization-gradient cooling of 1D and 2D ion Coulomb crystals}},\ }\href
	{https://doi.org/10.1088/1367-2630/abb912} {\bibfield  {journal} {\bibinfo
			{journal} {New J. Phys.}\ }\textbf {\bibinfo {volume} {22}},\ \bibinfo
		{pages} {103013} (\bibinfo {year} {2020})}\BibitemShut {NoStop}%
	\bibitem [{\citenamefont {Takahashi}\ \emph {et~al.}(2020)\citenamefont
		{Takahashi}, \citenamefont {Kassa}, \citenamefont {Christoforou},\ and\
		\citenamefont {Keller}}]{Takahashi2020}%
	\BibitemOpen
	\bibfield  {author} {\bibinfo {author} {\bibfnamefont {H.}~\bibnamefont
			{Takahashi}}, \bibinfo {author} {\bibfnamefont {E.}~\bibnamefont {Kassa}},
		\bibinfo {author} {\bibfnamefont {C.}~\bibnamefont {Christoforou}},\ and\
		\bibinfo {author} {\bibfnamefont {M.}~\bibnamefont {Keller}},\ }\bibfield
	{title} {\bibinfo {title} {{Strong Coupling of a Single Ion to an Optical
				Cavity}},\ }\href {https://doi.org/10.1103/PhysRevLett.124.013602} {\bibfield
		{journal} {\bibinfo  {journal} {Phys. Rev. Lett.}\ }\textbf {\bibinfo
			{volume} {124}},\ \bibinfo {pages} {013602} (\bibinfo {year}
		{2020})}\BibitemShut {NoStop}%
	\bibitem [{\citenamefont {Stute}\ \emph {et~al.}(2013)\citenamefont {Stute},
		\citenamefont {Casabone}, \citenamefont {Brandst{\"{a}}tter}, \citenamefont
		{Friebe}, \citenamefont {Northup},\ and\ \citenamefont {Blatt}}]{Stute2013}%
	\BibitemOpen
	\bibfield  {author} {\bibinfo {author} {\bibfnamefont {A.}~\bibnamefont
			{Stute}}, \bibinfo {author} {\bibfnamefont {B.}~\bibnamefont {Casabone}},
		\bibinfo {author} {\bibfnamefont {B.}~\bibnamefont {Brandst{\"{a}}tter}},
		\bibinfo {author} {\bibfnamefont {K.}~\bibnamefont {Friebe}}, \bibinfo
		{author} {\bibfnamefont {T.~E.}\ \bibnamefont {Northup}},\ and\ \bibinfo
		{author} {\bibfnamefont {R.}~\bibnamefont {Blatt}},\ }\bibfield  {title}
	{\bibinfo {title} {{Quantum-state transfer from an ion to a photon}},\ }\href
	{https://doi.org/10.1038/nphoton.2012.358} {\bibfield  {journal} {\bibinfo
			{journal} {Nat. Photonics}\ }\textbf {\bibinfo {volume} {7}},\ \bibinfo
		{pages} {219} (\bibinfo {year} {2013})}\BibitemShut {NoStop}%
	\bibitem [{\citenamefont {Walker}\ \emph {et~al.}(2020)\citenamefont {Walker},
		\citenamefont {Kashanian}, \citenamefont {Ward},\ and\ \citenamefont
		{Keller}}]{Walker2020}%
	\BibitemOpen
	\bibfield  {author} {\bibinfo {author} {\bibfnamefont {T.}~\bibnamefont
			{Walker}}, \bibinfo {author} {\bibfnamefont {S.~V.}\ \bibnamefont
			{Kashanian}}, \bibinfo {author} {\bibfnamefont {T.}~\bibnamefont {Ward}},\
		and\ \bibinfo {author} {\bibfnamefont {M.}~\bibnamefont {Keller}},\
	}\bibfield  {title} {\bibinfo {title} {{Improving the indistinguishability of
				single photons from an ion-cavity system}},\ }\href
	{https://doi.org/10.1103/PhysRevA.102.032616} {\bibfield  {journal} {\bibinfo
			{journal} {Phys. Rev. A}\ }\textbf {\bibinfo {volume} {102}},\ \bibinfo
		{pages} {032616} (\bibinfo {year} {2020})}\BibitemShut {NoStop}%
	\bibitem [{\citenamefont {Rempe}\ \emph {et~al.}(1992)\citenamefont {Rempe},
		\citenamefont {Lalezari}, \citenamefont {Thompson},\ and\ \citenamefont
		{Kimble}}]{Rempe1992}%
	\BibitemOpen
	\bibfield  {author} {\bibinfo {author} {\bibfnamefont {G.}~\bibnamefont
			{Rempe}}, \bibinfo {author} {\bibfnamefont {R.}~\bibnamefont {Lalezari}},
		\bibinfo {author} {\bibfnamefont {R.~J.}\ \bibnamefont {Thompson}},\ and\
		\bibinfo {author} {\bibfnamefont {H.~J.}\ \bibnamefont {Kimble}},\ }\bibfield
	{title} {\bibinfo {title} {{Measurement of ultralow losses in an optical
				interferometer}},\ }\href {https://doi.org/10.1364/OL.17.000363} {\bibfield
		{journal} {\bibinfo  {journal} {Opt. Lett.}\ }\textbf {\bibinfo {volume}
			{17}},\ \bibinfo {pages} {363} (\bibinfo {year} {1992})}\BibitemShut
	{NoStop}%
	\bibitem [{\citenamefont {Nguyen}\ \emph {et~al.}(2018)\citenamefont {Nguyen},
		\citenamefont {Utama}, \citenamefont {Lewty},\ and\ \citenamefont
		{Kurtsiefer}}]{Nguyen2018}%
	\BibitemOpen
	\bibfield  {author} {\bibinfo {author} {\bibfnamefont {C.~H.}\ \bibnamefont
			{Nguyen}}, \bibinfo {author} {\bibfnamefont {A.~N.}\ \bibnamefont {Utama}},
		\bibinfo {author} {\bibfnamefont {N.}~\bibnamefont {Lewty}},\ and\ \bibinfo
		{author} {\bibfnamefont {C.}~\bibnamefont {Kurtsiefer}},\ }\bibfield  {title}
	{\bibinfo {title} {{Operating a near-concentric cavity at the last stable
				resonance}},\ }\href {https://doi.org/10.1103/PhysRevA.98.063833} {\bibfield
		{journal} {\bibinfo  {journal} {Phys. Rev. A}\ }\textbf {\bibinfo {volume}
			{98}},\ \bibinfo {pages} {063833} (\bibinfo {year} {2018})}\BibitemShut
	{NoStop}%
	\bibitem [{\citenamefont {Casabone}\ \emph {et~al.}(2015)\citenamefont
		{Casabone}, \citenamefont {Friebe}, \citenamefont {Brandst{\"{a}}tter},
		\citenamefont {Sch{\"{u}}ppert}, \citenamefont {Blatt},\ and\ \citenamefont
		{Northup}}]{Casabone2015}%
	\BibitemOpen
	\bibfield  {author} {\bibinfo {author} {\bibfnamefont {B.}~\bibnamefont
			{Casabone}}, \bibinfo {author} {\bibfnamefont {K.}~\bibnamefont {Friebe}},
		\bibinfo {author} {\bibfnamefont {B.}~\bibnamefont {Brandst{\"{a}}tter}},
		\bibinfo {author} {\bibfnamefont {K.}~\bibnamefont {Sch{\"{u}}ppert}},
		\bibinfo {author} {\bibfnamefont {R.}~\bibnamefont {Blatt}},\ and\ \bibinfo
		{author} {\bibfnamefont {T.~E.}\ \bibnamefont {Northup}},\ }\bibfield
	{title} {\bibinfo {title} {{Enhanced Quantum Interface with Collective
				Ion-Cavity Coupling}},\ }\href
	{https://doi.org/10.1103/PhysRevLett.114.023602} {\bibfield  {journal}
		{\bibinfo  {journal} {Phys. Rev. Lett.}\ }\textbf {\bibinfo {volume} {114}},\
		\bibinfo {pages} {023602} (\bibinfo {year} {2015})}\BibitemShut {NoStop}%
	\bibitem [{\citenamefont {Cetina}\ \emph {et~al.}(2013)\citenamefont {Cetina},
		\citenamefont {Bylinskii}, \citenamefont {Karpa}, \citenamefont {Gangloff},
		\citenamefont {Beck}, \citenamefont {Ge}, \citenamefont {Scholz},
		\citenamefont {Grier}, \citenamefont {Chuang},\ and\ \citenamefont
		{Vuleti{\'{c}}}}]{Cetina2013}%
	\BibitemOpen
	\bibfield  {author} {\bibinfo {author} {\bibfnamefont {M.}~\bibnamefont
			{Cetina}}, \bibinfo {author} {\bibfnamefont {A.}~\bibnamefont {Bylinskii}},
		\bibinfo {author} {\bibfnamefont {L.}~\bibnamefont {Karpa}}, \bibinfo
		{author} {\bibfnamefont {D.}~\bibnamefont {Gangloff}}, \bibinfo {author}
		{\bibfnamefont {K.~M.}\ \bibnamefont {Beck}}, \bibinfo {author}
		{\bibfnamefont {Y.}~\bibnamefont {Ge}}, \bibinfo {author} {\bibfnamefont
			{M.}~\bibnamefont {Scholz}}, \bibinfo {author} {\bibfnamefont {A.~T.}\
			\bibnamefont {Grier}}, \bibinfo {author} {\bibfnamefont {I.}~\bibnamefont
			{Chuang}},\ and\ \bibinfo {author} {\bibfnamefont {V.}~\bibnamefont
			{Vuleti{\'{c}}}},\ }\bibfield  {title} {\bibinfo {title} {{One-dimensional
				array of ion chains coupled to an optical cavity}},\ }\href
	{https://doi.org/10.1088/1367-2630/15/5/053001} {\bibfield  {journal}
		{\bibinfo  {journal} {New J. Phys.}\ }\textbf {\bibinfo {volume} {15}},\
		\bibinfo {pages} {053001} (\bibinfo {year} {2013})}\BibitemShut {NoStop}%
	\bibitem [{\citenamefont {Begley}\ \emph {et~al.}(2016)\citenamefont {Begley},
		\citenamefont {Vogt}, \citenamefont {Gulati}, \citenamefont {Takahashi},\
		and\ \citenamefont {Keller}}]{Begley2016}%
	\BibitemOpen
	\bibfield  {author} {\bibinfo {author} {\bibfnamefont {S.}~\bibnamefont
			{Begley}}, \bibinfo {author} {\bibfnamefont {M.}~\bibnamefont {Vogt}},
		\bibinfo {author} {\bibfnamefont {G.~K.}\ \bibnamefont {Gulati}}, \bibinfo
		{author} {\bibfnamefont {H.}~\bibnamefont {Takahashi}},\ and\ \bibinfo
		{author} {\bibfnamefont {M.}~\bibnamefont {Keller}},\ }\bibfield  {title}
	{\bibinfo {title} {{Optimized Multi-Ion Cavity Coupling}},\ }\href
	{https://doi.org/10.1103/PhysRevLett.116.223001} {\bibfield  {journal}
		{\bibinfo  {journal} {Phys. Rev. Lett.}\ }\textbf {\bibinfo {volume} {116}},\
		\bibinfo {pages} {223001} (\bibinfo {year} {2016})}\BibitemShut {NoStop}%
	\bibitem [{\citenamefont {Lamata}\ \emph {et~al.}(2011)\citenamefont {Lamata},
		\citenamefont {Leibrandt}, \citenamefont {Chuang}, \citenamefont {Cirac},
		\citenamefont {Lukin}, \citenamefont {Vuleti{\'{c}}},\ and\ \citenamefont
		{Yelin}}]{Lamata2011}%
	\BibitemOpen
	\bibfield  {author} {\bibinfo {author} {\bibfnamefont {L.}~\bibnamefont
			{Lamata}}, \bibinfo {author} {\bibfnamefont {D.~R.}\ \bibnamefont
			{Leibrandt}}, \bibinfo {author} {\bibfnamefont {I.~L.}\ \bibnamefont
			{Chuang}}, \bibinfo {author} {\bibfnamefont {J.~I.}\ \bibnamefont {Cirac}},
		\bibinfo {author} {\bibfnamefont {M.~D.}\ \bibnamefont {Lukin}}, \bibinfo
		{author} {\bibfnamefont {V.}~\bibnamefont {Vuleti{\'{c}}}},\ and\ \bibinfo
		{author} {\bibfnamefont {S.~F.}\ \bibnamefont {Yelin}},\ }\bibfield  {title}
	{\bibinfo {title} {{Ion Crystal Transducer for Strong Coupling between Single
				Ions and Single Photons}},\ }\href
	{https://doi.org/10.1103/PhysRevLett.107.030501} {\bibfield  {journal}
		{\bibinfo  {journal} {Phys. Rev. Lett.}\ }\textbf {\bibinfo {volume} {107}},\
		\bibinfo {pages} {030501} (\bibinfo {year} {2011})}\BibitemShut {NoStop}%
	\bibitem [{\citenamefont {Gerritsma}\ \emph {et~al.}(2008)\citenamefont
		{Gerritsma}, \citenamefont {Kirchmair}, \citenamefont {Z{\"{a}}hringer},
		\citenamefont {Benhelm}, \citenamefont {Blatt},\ and\ \citenamefont
		{Roos}}]{Gerritsma2008}%
	\BibitemOpen
	\bibfield  {author} {\bibinfo {author} {\bibfnamefont {R.}~\bibnamefont
			{Gerritsma}}, \bibinfo {author} {\bibfnamefont {G.}~\bibnamefont
			{Kirchmair}}, \bibinfo {author} {\bibfnamefont {F.}~\bibnamefont
			{Z{\"{a}}hringer}}, \bibinfo {author} {\bibfnamefont {J.}~\bibnamefont
			{Benhelm}}, \bibinfo {author} {\bibfnamefont {R.}~\bibnamefont {Blatt}},\
		and\ \bibinfo {author} {\bibfnamefont {C.~F.}\ \bibnamefont {Roos}},\
	}\bibfield  {title} {\bibinfo {title} {{Precision measurement of the
				branching fractions of the 4p 2P3/2 decay of Ca II}},\ }\href
	{https://doi.org/10.1140/epjd/e2008-00196-9} {\bibfield  {journal} {\bibinfo
			{journal} {Eur. Phys. J. D}\ }\textbf {\bibinfo {volume} {50}},\ \bibinfo
		{pages} {13} (\bibinfo {year} {2008})}\BibitemShut {NoStop}%
	\bibitem [{\citenamefont {Siegman}(1986)}]{siegman86}%
	\BibitemOpen
	\bibfield  {author} {\bibinfo {author} {\bibnamefont {Siegman}},\ }\href@noop
	{} {\emph {\bibinfo {title} {{Lasers}}}}\ (\bibinfo  {publisher} {University
		Science Books},\ \bibinfo {year} {1986})\ pp.\ \bibinfo {pages}
	{744--776}\BibitemShut {NoStop}%
	\bibitem [{\citenamefont {Hood}\ \emph {et~al.}(2001)\citenamefont {Hood},
		\citenamefont {Kimble},\ and\ \citenamefont {Ye}}]{Hood2001}%
	\BibitemOpen
	\bibfield  {author} {\bibinfo {author} {\bibfnamefont {C.~J.}\ \bibnamefont
			{Hood}}, \bibinfo {author} {\bibfnamefont {H.~J.}\ \bibnamefont {Kimble}},\
		and\ \bibinfo {author} {\bibfnamefont {J.}~\bibnamefont {Ye}},\ }\bibfield
	{title} {\bibinfo {title} {{Characterization of high-finesse mirrors: Loss,
				phase shifts, and mode structure in an optical cavity}},\ }\href
	{https://doi.org/10.1103/PhysRevA.64.033804} {\bibfield  {journal} {\bibinfo
			{journal} {Phys. Rev. A}\ }\textbf {\bibinfo {volume} {64}},\ \bibinfo
		{pages} {033804} (\bibinfo {year} {2001})}\BibitemShut {NoStop}%
	\bibitem [{\citenamefont {Black}(2001)}]{Black2001}%
	\BibitemOpen
	\bibfield  {author} {\bibinfo {author} {\bibfnamefont {E.~D.}\ \bibnamefont
			{Black}},\ }\bibfield  {title} {\bibinfo {title} {{An introduction to
				Pound–Drever–Hall laser frequency stabilization}},\ }\href
	{https://doi.org/10.1119/1.1286663} {\bibfield  {journal} {\bibinfo
			{journal} {Am. J. Phys}\ }\textbf {\bibinfo {volume} {69}},\ \bibinfo {pages}
		{79} (\bibinfo {year} {2001})}\BibitemShut {NoStop}%
	\bibitem [{\citenamefont {Roos}(2000)}]{Roos}%
	\BibitemOpen
	\bibfield  {author} {\bibinfo {author} {\bibfnamefont {C.}~\bibnamefont
			{Roos}},\ }\emph {\bibinfo {title} {{Controlling the quantum state of trapped
				ions}}},\ \href@noop {} {\bibinfo {type} {Phd thesis}},\ \bibinfo  {school}
	{University of Innsbruck} (\bibinfo {year} {2000})\BibitemShut {NoStop}%
	\bibitem [{\citenamefont {Wineland}\ \emph {et~al.}(1998)\citenamefont
		{Wineland}, \citenamefont {Monroe}, \citenamefont {Itano}, \citenamefont
		{Leibfried}, \citenamefont {King},\ and\ \citenamefont
		{Meekhof}}]{Wineland1998}%
	\BibitemOpen
	\bibfield  {author} {\bibinfo {author} {\bibfnamefont {D.}~\bibnamefont
			{Wineland}}, \bibinfo {author} {\bibfnamefont {C.}~\bibnamefont {Monroe}},
		\bibinfo {author} {\bibfnamefont {W.}~\bibnamefont {Itano}}, \bibinfo
		{author} {\bibfnamefont {D.}~\bibnamefont {Leibfried}}, \bibinfo {author}
		{\bibfnamefont {B.}~\bibnamefont {King}},\ and\ \bibinfo {author}
		{\bibfnamefont {D.}~\bibnamefont {Meekhof}},\ }\bibfield  {title} {\bibinfo
		{title} {{Experimental issues in coherent quantum-state manipulation of
				trapped atomic ions}},\ }\href {https://doi.org/10.6028/jres.103.019}
	{\bibfield  {journal} {\bibinfo  {journal} {J. Res. Nat. Inst. Stand.
				Technol.}\ }\textbf {\bibinfo {volume} {103}},\ \bibinfo {pages} {259}
		(\bibinfo {year} {1998})}\BibitemShut {NoStop}%
	\bibitem [{\citenamefont {Je{\v{z}}ek}\ \emph {et~al.}(2003)\citenamefont
		{Je{\v{z}}ek}, \citenamefont {Fiur{\'{a}}{\v{s}}ek},\ and\ \citenamefont
		{Hradil}}]{Jezek2003}%
	\BibitemOpen
	\bibfield  {author} {\bibinfo {author} {\bibfnamefont {M.}~\bibnamefont
			{Je{\v{z}}ek}}, \bibinfo {author} {\bibfnamefont {J.}~\bibnamefont
			{Fiur{\'{a}}{\v{s}}ek}},\ and\ \bibinfo {author} {\bibfnamefont
			{Z.}~\bibnamefont {Hradil}},\ }\bibfield  {title} {\bibinfo {title} {{Quantum
				inference of states and processes}},\ }\href
	{https://doi.org/10.1103/PhysRevA.68.012305} {\bibfield  {journal} {\bibinfo
			{journal} {Phys. Rev. A}\ }\textbf {\bibinfo {volume} {68}},\ \bibinfo
		{pages} {012305} (\bibinfo {year} {2003})}\BibitemShut {NoStop}%
\end{thebibliography}

%

\end{document}